\documentclass[prd,aps,preprint,tightenlines,superscriptaddress]{revtex4}
\usepackage{epsfig}
\usepackage{axodraw}
\usepackage{amsmath}
\begin{document}

\preprint{DOE/ER/40762-308 \ UM-PP\#04-037}

\title{QCD Factorization for Semi-Inclusive Deep-Inelastic Scattering \\ at Low Transverse Momentum}
\author{Xiangdong Ji}
\affiliation{Department of Physics,
University of Maryland,
College Park, Maryland 20742, USA}
\author{Jian-Ping Ma}
\affiliation{Institute of Theoretical Physics, Academia Sinica,
Beijing, 100080, P. R. China} \affiliation{Department of Physics,
University of Maryland, College Park, Maryland 20742, USA}
\author{Feng Yuan}
\affiliation{Department of Physics,
University of Maryland,
College Park, Maryland 20742, USA}

\date{\today}
\vspace{0.5in}
\begin{abstract}
We demonstrate a factorization formula for semi-inclusive
deep-inelastic scattering with hadrons in the current
fragmentation region detected at low transverse momentum. To
facilitate the factorization, we introduce the transverse-momentum
dependent parton distributions and fragmentation functions with
gauge links slightly off the light-cone, and with soft-gluon
radiations subtracted. We verify the factorization to one-loop
order in perturbative quantum chromodynamics and argue that it is
valid to all orders in perturbation theory.

\end{abstract}

\maketitle

\section{Introduction}

In recent years, semi-inclusive deep-inelastic (SIDIS)
lepton-nucleon scattering has emerged as an important tool to
learn various aspects of perturbative and non-perturbative quantum
chromodynamics (pQCD), the internal structure of the nucleon, in
particular. The European Muon Collaboration experiment at CERN has
provided us valuable information about the flavor dependence of
quark fragmentation functions \cite{EMC89}. The H1 and ZEUS
collaborations at the DESY HERA collider have measured the
topology of the hadron final states in great detail and have
compared them with the predictions of perturbative QCD\cite{HERA}.
In the area of polarized semi-inclusive DIS, the Spin Muon
Collaboration, and recently the HERMES collaboration at DESY, have
extracted the sea quark distributions and the polarized gluon
distribution with controlled accuracy \cite{SMC,HERMES}. More
recently, the target single-spin asymmetry measured by HERMES in
semi-inclusive DIS is a new observable sensitive, for example, to
the quark transversity distribution through the transverse
momentum dependence of the produced hadron \cite{SSA}.

In the semi-inclusive production of DIS, both the longitudinal
momentum fraction $z$ and the transverse-momentum $P_{h\perp}$ of
the hadron yield can be measured. When the transverse momentum is
integrated over or when it is comparable to the hard photon-mass
scale, $P_{h\perp}\sim Q$, the cross sections can be calculated
from the standard pQCD formalism similar to inclusive DIS and
Feynman parton distributions \cite{MenOlnSop92}. In these cases,
the theoretical tool has been well tested against experimental
data with notable successes. When the transverse-momentum is much
smaller than $Q$, but is still hard, $P_{h\perp}\gg\Lambda_{\rm
QCD}$, the cross section can be calculated again with integrated
parton distributions augmented by small non-perturbative QCD
corrections. The hard part contains the large double logarithms of
the type $\alpha_s\ln^2P_{h\perp}/Q^2$. To make reliable
predictions, these large logarithms must be summed
\cite{DokDyaTro80,ParPet79}. An adequate formalism was developed
by Collins and Soper in the case of $e^+e^-$ annihilation
\cite{ColSop81}, and shortly thereafter applied to the Drell-Yan
process by Collins, Soper and Sterman (CSS) \cite{ColSopSte85}. A
first application of the CSS approach to SIDIS was made by Meng,
Olness, and Soper \cite{MenOlnSop96}. Recently, a quantitative
comparison between this theory and data from HERA collider has
been made by Nadolsky, Stump and Yuan \cite{NadStuYua99}.

In this paper, we are interested in a special kinematic regime in
SIDIS where $P_{h\perp}$ is soft, i.e. on the order of
$\Lambda_{\rm QCD}$, and $Q^2$ is not too large, for example, on
the order of tens or hundreds of GeV$^2$. When $Q^2$ is large, the
soft gluon radiations become important and can easily generate a
large transverse momentum $\gg \Lambda_{\rm QCD}$. Then the cross
section for the hadron yield with $P_{h\perp}\sim \Lambda_{\rm
QCD}$ is exponentially suppressed. To have a significant fraction
of events with $P_{h\perp}\sim \Lambda_{\rm QCD}$, fixed-target
experiments with lepton beam energies on the order of tens to
hundreds of GeV are preferred. The above kinematic regime is in
fact ideal for studying transverse-momentum dependent (TMD) parton
distributions in the nucleon and the related quark fragmentation
functions. Recent interest in this subject has been stimulated by
Collins's observation that semi-inclusive DIS at low-$P_{\perp}$
provides a tool to measure the quark transversity distribution
\cite{Col93}. The physics potential has been reinforced by the
rediscovery of Siver's effect \cite{Siv90} by Brodsky, Hwang, and
Schimdt \cite{BroHwaSch02}.

The main result of this paper is a QCD factorization theorem for
the SIDIS cross section in the above kinematics region, accurate
up to the power corrections $(P_{h\perp}^2/Q^2)^n$ and to all
orders in perturbation theory. This factorization has been
conjectured by Collins \cite{Col93} (Eq. (13)), following the
early work of Collins and Soper on $e^+e^-$ annihilation
\cite{ColSop81}. However, an exact statement of the factorization
theorem requires an adequate definition of the TMD parton
distributions and fragmentation functions in QCD and a systematic
factorization (and subtraction) of soft, collinear, and hard gluon
contributions. In light of the recent development in this area
\cite{Col89,ColHau00,BelJiYua03,Col03}, here we provide a first
detailed examination of QCD radiative corrections in SIDIS,
following the methodology of Ref. \cite{ColSop81}.

The factorization theorem we propose for the leading
spin-independent structure function is
\begin{eqnarray}
 F(x_B,z_h,P_{h\perp},Q^2)&=&\sum_{q=u,d,s,...} e_q^2\int d^2\vec{k}_\perp d^2\vec{p}_{\perp}
      d^2\vec{\ell}_\perp
   \nonumber \\
   && \times  q\left({x_B}, k_\perp,\mu^2,x_B\zeta, \rho\right)
    \hat q_T\left({z_h}, p_{\perp},\mu^2,\hat\zeta/z_h, \rho\right)  S(\vec{\ell}_\perp,\mu^2,\rho) \nonumber \\
&& \times H\left(Q^2,\mu^2,\rho\right)
\delta^2(z_h\vec{k}_\perp+\vec{p}_\perp +\vec{\ell}_\perp-
\vec{P}_{h\perp}) \ ,
\end{eqnarray}
where $\mu$ is a renormalization (and collinear factorization)
scale; $\rho$ is a gluon rapidity cut-off parameter; {\it the
$\mu$ and $\rho$ dependence cancels among various factors.} In a
special system of coordinates in which $x_B\zeta=\hat \zeta/z_h$,
one has $\zeta^2x_B^2 = \hat \zeta^2/z^2_h = Q^2\rho$. The
physical interpretation of the factors are as follows: $q$ is TMD
quark distribution function depending on, among others, the
Bjorken $x_B$; $\hat q$ is the TMD quark fragmentation function
depending on, among others, the hadron momentum fraction $z_h$;
$H$ represents the contribution of parton hard scattering and is a
perturbation series in $\alpha_s$; and, finally, the soft factor
$S$ comes from soft gluon radiations and is defined by a matrix
element of Wilson lines in QCD vacuum. The above result shows that
the hadron transverse momentum is generated from the combined
effects of transverse-momentum of the quarks in the nucleon, soft
gluon radiation, and the transverse-momentum of the quark
fragmentation. There is no direct contribution from the TMD gluon
distribution in this kinematic region.

The main steps to establish the above factorization is as follows.
In Sec. II, we introduce the TMD parton distribution and
fragmentation function, and calculate them to one-loop order in
perturbative QCD. The result contains both soft and collinear
divergences and obeys the Collins and Soper evolution equation in
the rapidity cut-off. We study the factorization of the TMD
distributions by subtracting away the soft contributions. In Sec.
III, one-loop result for semi-inclusive DIS scattering is
obtained, and the factorization is shown to be true on the
diagram-by-diagram basis. In Sec. IV, we generalize the one-loop
result to all orders by identifying the leading regions for an
arbitrary Feynman diagram using soft and collinear power counting.
We then argue that a systematic factorization of the leading
region leads to the general formula in Eq. (1). In Sec. V, the
large logarithms in the perturbative expression are summed through
solving evolution equations. We conclude the paper in Sec. VI.

The factorization considered here can also be studied in the
framework of soft-collinear effective theory developed recently in
Refs. \cite{BauFlePirSte01,BauSte01,BauPirSte02,BauEtc02}. We will
leave this subject for a future publication.

\section{Transverse-Momentum Dependent Parton Distribution}

In the factorization formula (Eq. (1)), there is a factor
$q(x,k_\perp, \mu^2, x_B\zeta, \rho)$ representing a
transverse-momentum dependent (TMD) parton distribution, which
differs from the usual Feynman parton distribution where the
parton transverse-momentum has already been integrated over. This
object was introduced by Collins and Soper in the axial gauge and
has a number of interesting properties \cite{ColSop81}. In
particular, it has a light-cone singularity and hence depends on
the energy of the parent nucleon (related to $\zeta$) in addition
to the parton's longitudinal momentum fraction $x$ and transverse
momentum $k_\perp$. The sensitivity to the small-$x$ gluon physics
is controlled by a parameter $\rho$.

In this paper, we follow a definition of TMD distribution in
Feynman gauge with explicit gauge links \cite{Col89}. We avoid the
axial gauge because of the potential existence of gauge links at
space-time infinity \cite{BelJiYua03}. We calculate the TMDPD at
one-loop order and show that it obeys the Collins-Soper evolution
equation. The simplest definition of the distribution contains the
soft-gluon effect which must be subtracted; we show how this can
be done at the one-loop level. We also discuss theoretical
difficulty to recover the integrated parton distribution from a
direct transverse-momentum integration.

\subsection{Definition of TMDPD}

Consider a hadron, a nucleon for example, with four-momentum $P$.
For convenience, we choose $\vec{P}$ along the $z$-direction,
$P^\mu= (P^0, 0, 0, P^3)$. In the limit $P^3\rightarrow \infty$,
the $P^\mu $ is proportional to the light-cone vector $(1,0,0,1)$.
From now on, we use the light-cone coordinates $k^\pm = (k^0\pm
k^3)/\sqrt{2}$, and write any four-vector $k^\mu$ in the form of
$(k^-,\vec{k})= (k^-, k^+, \vec{k}_\perp)$, where $\vec{k}_\perp$
represents two perpendicular components $(k^x,k^y)$. Let $(xP^+,
\vec{k}_\perp)$ represent the momentum of a parton (quark or
gluon) in the hadron. Let us start with the following definition
of the transverse-momentum dependent quark distribution in a class
of non-singular gauges \cite{Col89,Col03},
\begin{eqnarray}
       {\cal Q}(x, k_\perp, \mu, x\zeta)
        &=& \frac{1}{2}\int
        \frac{d\xi^-}{2\pi}e^{-ix\xi^-P^+}\int
        \frac{d^2\vec{b}_\perp}{(2\pi)^2} e^{i\vec{b}_\perp\cdot
        \vec{k}_\perp} \nonumber \\
   &&    \times \left\langle P\left|\overline{\psi}_q(\xi^-,0,\vec{b}_\perp){\cal
        L}^\dagger_{v}(\infty;\xi^-,0,\vec{b}_\perp) \gamma^+ {\cal
        L}_{v}(\infty;0)
        \psi_q(0)\right|P\right\rangle\ ,
        \label{tmdpd}
\end{eqnarray}
where the quark color indices are implicit, $\psi_q$ is the quark
field, $v^\mu$ is a time-like dimensionless ($v^2>0$) four-vector
with zero transverse components $(v^-,v^+,\vec{0})$, and ${\cal
L}_v$ is a gauge link along $v^\mu$,
\begin{equation}
    {\cal L}_{v}(\infty;\xi) = \exp\left(-ig\int^\infty_0 d\lambda v\cdot A(\lambda
    v +\xi)\right) \ .
\end{equation}
The sign convention for the gauge coupling is $D^\mu =\partial^\mu
+ igA^\mu$. As mentioned before, $\mu$ is an ultraviolet (UV)
renormalization (or cut-off) scale.

It is convenient to introduce the ``gauge-invariant" quark field,
\begin{equation}
   {\mit \Psi}_v(\xi) =   {\cal L}_{v}(\infty;\xi)\psi(\xi) \ .
\end{equation}
The quark distribution becomes simply
\begin{eqnarray}
      {\cal  Q}(x, k_\perp, \mu, x\zeta)
        = \frac{1}{2}\int
        \frac{d\xi^-}{2\pi}e^{-ix\xi^-P^+}\int
        \frac{d^2b_\perp}{(2\pi)^2} e^{i\vec{b}_\perp\cdot
        \vec{k}_\perp}  \left\langle P\left|\overline{\mit \Psi}_v(\xi^-,0,\vec{b}_\perp) \gamma^+
        {\mit \Psi}_v(0)\right|P\right\rangle\ .
\end{eqnarray}
The variable $\zeta^2$ denotes the combination $(2P\cdot
v)^2/v^2=\zeta^2$.

Physically, a parton interpretation of ${\cal Q}(x, k_\perp, \mu,
\zeta)$ is the most natural if $v$ is chosen along the conjugating
light-cone direction of $P^\mu$, i.e., $v^\mu \sim (1,0,0_\perp)$.
However, as has been known in the literature and re-emphasized
recently by Collins \cite{Col03}, the distribution in this limit
has logarithmic divergences (also called light-cone singularity)
corresponding to contributions of virtual gluons with zero plus
momentum $\ell^+$, or infinitely negative rapidity, $\ln
\ell^+/\ell^-$. [In a physical process, the plus momentum of a
parton is limited by the kinematics of scattering.] To avoid the
divergence, we must introduce a rapidity cut-off for the gluons.
One way to accomplish this is to introduce a non-light-like
$v^\mu$, such as with a $v^+\ne 0$ \cite{Col89}. Then the
contribution of the virtual gluons with rapidity smaller than $\ln
v^+/v^-$ is excluded from the parton distribution. As a
consequence, a dimensional scalar $\zeta^2 = (2P\cdot v)^2/v^2$
emerges in the distribution. The limit of lifting the cut-off,
$v^+\rightarrow 0$, corresponds to $\zeta\rightarrow \infty$. [In
the following expressions, we will take this limit whenever we
can.] The light-cone divergences are now reflected in the large
logarithms involving $\zeta$. The $\zeta$-evolution of the TMD
distribution can be viewed as either the evolution in the gluon
rapidity cut-off through $v^\mu$ or that in the energy of the
incoming hadron. The evolution is calculable in perturbation
theory when $k_\perp$ is hard, i.e., $\gg \Lambda_{\rm QCD}$
\cite{ColSop81}.

Unless stated otherwise, we work in non-singular gauges, such as
covariant gauges (including the Feynman gauge), for which the
gauge potential vanishes at space-time infinity. In light-cone
gauge, however, it is well known that the gauge potential is
finite at infinity. Otherwise, the single spin asymmetry discussed
by Brodsky {\it et al.} would disappear in such a gauge
\cite{BroHwaSch02}. Belitsky, Ji and Yuan have shown that in
singular gauges, one generally has to include gauge links at
infinity \cite{BelJiYua03}. These extra links can be found by
imposing the gauge invariance of the parton densities starting
from their definition in non-singular gauges.

Since the two quark fields in Eq. (\ref{tmdpd}) are separated
along the spatial directions, the only ultraviolet divergence in
${\cal Q}(x,k_\perp, \mu, \zeta)$ comes from the wave function
renormalization of the quark fields and the gauge links. In this
paper, we use dimensional regularization (DR) and modified minimal
subtraction ($\overline{{\rm MS}}$) to treat ultraviolet
divergences. If we use Eq. (\ref{tmdpd}) naively in the axial
gauge $v\cdot A=0$, then the ultraviolet divergence of ${\cal
Q}(x,k_\perp)$ is the same as the quark wave function
renormalization in that gauge. Then the renormalization group
equation becomes simple,
\begin{equation}
       \mu \frac{d{\cal Q}(x,k_\perp, \mu,x\zeta)}{d\mu}
        = 2\gamma_F {\cal Q}(x,k_\perp,\mu, x\zeta) \ ,
\end{equation}
where $\gamma_F$ is the anomalous dimension of the quark field in
the axial gauge: $\gamma_F = (3\alpha_s/4\pi)C_F + {\cal
O}(\alpha_s^2)$.

\subsection{One-Loop Calculation}

In this subsection, we present the one-loop result for a quark TMD
distribution in an ``on-shell" quark. The calculation is important
for a number of reasons. First, it shows clearly that the TMD
distribution contains double logarithms in $\zeta$ because of the
collinear and light-cone divergences. It also serves as an
explicit check for the evolution equation in rapidity cut-off.
More importantly, the one-loop result allows one to devise QCD
factorizations both for the distribution itself and for one-loop
DIS cross section to be presented in the next section.

We use a non-zero gluon mass $\lambda$ as an infrared regulator
since there is no non-linear gluon coupling at one loop. One can
use dimensional regularization beyond the leading order. The
factorization is, of course, independent of the infrared
regulator. Collinear singularities are regulated by none-zero
quark masses.

\begin{figure}
\SetWidth{0.7}
\begin{center} \begin{picture}(330,70)(0,0)
\SetWidth{1.0}

\SetOffset(0,15)
\ArrowLine(0,0)(0,15)\ArrowLine(0,35)(0,50)\Line(0,15)(0,35)
\GlueArc(0,25)(10,-90,90){2.0}{5}
\ArrowLine(60,50)(60,0)\DashLine(35,55)(35,-5){5}
\Text(85,25)[c]{+}\Text(30,-15)[c]{(a)}

\SetOffset(110,15) \ArrowLine(0,0)(0,50)
\Line(0,50)(30,50)\Line(0,48)(30,48)\GlueArc(20,50)(10,180,360){2.0}{5}
\ArrowLine(60,50)(60,0)\DashLine(40,55)(40,-5){5}
\Text(85,25)[c]{+}\Text(30,-15)[c]{(b)}

 \SetOffset(220,15)
\ArrowLine(0,0)(0,20)\ArrowLine(0,20)(0,50)
\Line(0,50)(30,50)\Line(0,48)(30,48)\Gluon(0,20)(30,50){-2.0}{6}
\ArrowLine(60,50)(60,0)\DashLine(40,55)(40,-5){5}
 \Text(30,-15)[c]{(c)}
\end{picture}
\end{center}
\caption{\it Virtual-gluon contribution to one-loop
transverse-momentum dependent quark distribution in an on-shell
quark. The asymmetric diagrams from left-right reflection are not
shown, but are included in the result.}
\end{figure}
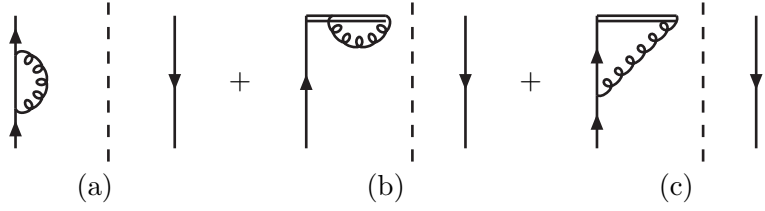

Let us first consider the virtual contribution shown in Fig. 1.
For the self-energy diagram on the incoming quark leg, one has a
contribution $
          {\cal Q}(x, k_\perp) = \delta(x-1)
           \delta^2(\vec{k}_\perp)(Z_F-1)$ with
\begin{equation}
     Z_F = 1+ \frac{\alpha_sC_F}{4\pi}\left(
     -\ln\frac{\mu^2}{m^2} + 2\ln\frac{m^2}{\lambda^2}-4\right) \
     ,
\label{zf}
\end{equation}
where $m$ and $\lambda$ are the masses of the quark and gluon,
respectively, $C_F=(N_c^2-1)/(2N_c)$ with $N_c=3$; a term linear
in $N_\epsilon = 2/\epsilon - \gamma_E + \ln 4\pi$, where
$\epsilon=4-d$ and $\gamma_E$ the Euler constant, has been removed
according to the ${\rm \overline{MS}}$ scheme. The on-shell
renormalization introduces the soft divergence in $Z_F$,
reflecting in the gluon mass dependence. For the self-energy on
the gauge link, one has a similar contribution with $Z_F$ replaced
by,
\begin{equation}
     Z_W =1 +\frac{C_F\alpha_s}{4\pi}\left(
     2\ln\frac{\mu^2}{\lambda^2}\right) \ .
\end{equation}
Finally, the diagram with the virtual gluon vertex again has a
similar contribution with $Z_F$ replaced by
\begin{eqnarray}
   Z_V  = 1 + \frac{\alpha_s C_F}{4\pi}\left[
     2\ln \frac{\mu^2}{m^2} + 2 \ln \frac{\zeta^2}{m^2}
 - \ln^2\frac{\zeta^2}{m^2}  -2 \ln \frac{m^2}{\lambda^2}\ln \frac{\zeta^2}{m^2}
 -\frac{2\pi^2}{3}+4 \right] \ ,
\end{eqnarray}
where we have made the approximation that $\zeta^2 =
\frac{4(P\cdot v)^2}{v^2}$ is much larger than any other soft QCD
scales. When $\zeta$ is large, the double logarithms slow down the
convergence of the pQCD series and call for a resummation which
can be accomplished with the Collins-Soper equation (see the next
subsection). In summary, the virtual diagrams give,
\begin{equation}
 {\cal Q}(x, k_\perp,\mu,x\zeta)|_{\rm fig.1} = \delta(x-1)
           \delta^2(\vec{k}_\perp)(Z_F+Z_W+Z_V-3) \ ,
\label{fig1}
\end{equation}
where the dependence on soft scales $m$ and $\lambda$ is implicit
on the left.

\begin{figure}
\begin{center} \begin{picture}(330,70)(0,0)
\SetWidth{1.0}

\SetOffset(0,15) \ArrowLine(0,0)(0,25)\ArrowLine(0,25)(0,50)
\Gluon(0,25)(60,25){-2.5}{8}
\ArrowLine(60,50)(60,25)\ArrowLine(60,25)(60,0)\DashLine(30,55)(30,-5){5}
\Text(85,25)[c]{+} \Text(30,-15)[c]{(a)}

\SetOffset(110,15) \ArrowLine(0,0)(0,25)\ArrowLine(0,25)(0,50)
\Gluon(0,25)(40,50){-2.5}{6}\Line(60,50)(40,50)\Line(60,48)(40,48)
\ArrowLine(60,50)(60,0)\DashLine(25,55)(25,-5){5}
\Text(85,25)[c]{+}\Text(30,-15)[c]{(b)}

 \SetOffset(220,15)\ArrowLine(0,0)(0,50)
\Line(0,50)(20,50)\Line(0,48)(20,48)\Line(60,50)(40,50)\Line(60,48)(40,48)
\GlueArc(30,50)(10,180,360){2.0}{5}
\ArrowLine(60,50)(60,0)\DashLine(30,55)(30,-5){5}
\Text(30,-15)[c]{(c)}
\end{picture}
\end{center}
\caption{\it Same as Fig. 1: real-gluon contribution.}
\end{figure}
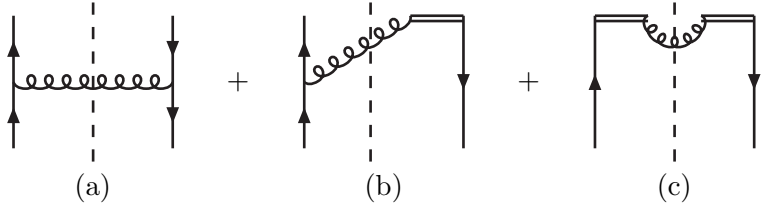

Now turn to the real gluon emission contributions shown in Fig. 2.
The contribution from the diagram 2a) without the light-cone link,
\begin{eqnarray}
   {\cal  Q}(x, k_\perp,\mu,x\zeta)|_{\rm fig.2a} &=& \frac{\alpha_s
    C_F|1-x|}{2\pi^2}\left[\frac{1}{k_\perp^2 + x\lambda^2 +
    (1-x)^2m^2} \right. \nonumber \\
  && \left.- \frac{2xm^2}{(k_\perp^2+ x\lambda^2 +
  (1-x)^2m^2)^2}\right]\ ,
\label{real-no-link}
\end{eqnarray}
where we have taken $\epsilon\rightarrow 0$. This must be done if
we treat the TMDPD in the factorization formula as a physical
observable. This, however, introduces certain problems in
integrating out $k_\perp$ in DR, and we will discuss this more
thoroughly in Sec. II.F. The transverse-momentum $k_\perp$ can go
to zero, and therefore we cannot set the quark and gluon masses to
zero too soon. However, the non-perturbative QCD physics will
erase this sensitivity after factorization is formulated.

The contribution from the diagram 2b), including the hermitian
conjugation term, is,
\begin{equation}
 {\cal Q}(x, k_\perp, \mu, x\zeta)|_{\rm fig.2b} = \frac{\alpha_s
    C_F}{\pi^2} \frac{x}{|1-x|}
    \left[\frac{1}{k_\perp^2 + x\lambda^2 + (1-x)^2m^2} -
    \frac{1}{k_\perp^2 + \lambda^2 + \zeta^2(1-x)^2}\right] \ . \label{real-one-link}
\end{equation}
The second term regularizes the light-cone singularity at $x=1$.
If one uses the usual regularization method of a plus function
\cite{AltPar77} and take the limit that $\zeta$ is large, the
above can be transformed into the following form,
\begin{eqnarray}
{\cal Q}(x, k_\perp, \mu, x\zeta) |_{\rm fig.2b} &=&
\frac{\alpha_s
    C_F}{\pi^2} \frac{x}{(1-x)_+}
    \frac{1}{k_\perp^2 + x\lambda^2 + (1-x)^2m^2} \nonumber \\
   && + \frac{\alpha_s
    C_F}{2\pi^2}\delta(x-1)\frac{1}{k_\perp^2+\lambda^2}\ln\frac{\zeta^2}{k_\perp^2+\lambda^2}
    \ ,
\end{eqnarray}
which has a delta function at $x=1$.

Finally the contribution from the diagram 2c) with two gauge links
is
\begin{equation}
 {\cal Q}(x, k_\perp,\mu,x\zeta)|_{\rm fig.2c} = -\frac{\alpha_s C_F}{2\pi^2}
 \frac{2|1-x|\zeta^2}{\left(k_\perp^2 + \lambda^2 + (1-x)^2
 \zeta^2\right)^2}\ .
 \label{real-two-link}
\end{equation}
Formally, it vanishes when $\zeta$ is large, except in the region
where $(1-x)^2\zeta^2$ is small. Therefore in the limit
$\zeta\rightarrow \infty$, the above is the same as a $\delta$
function at $x=1$,
\begin{equation}
   {\cal Q}(x, k_\perp, \mu, x\zeta)|_{\rm fig.2c} =- \delta(x-1)
 \frac{\alpha_s}{2\pi^2}C_F \frac{1}{k_\perp^2+\lambda^2}  \ .
\end{equation}
We caution the reader that taking $\zeta\rightarrow \infty$ limit
conflicts with the $k_\perp\rightarrow \infty$ limit. In fact, in
the above example, it turns a $k_\perp$-convergent integral into a
divergent one.

\subsection{Collins-Soper Evolution in Hadron Energy or Gluon Rapidity Cut-Off}

As we have seen from the previous subsection, unlike the Feynman
parton distributions which contain just the collinear
singularities from the quark masses, the TMD distributions contain
in addition the light-cone singularities which are regulated by
$\zeta^2$. Since $\zeta\rightarrow \infty$ corresponds to the
high-energy limit, $\zeta$ dependence of the parton distribution
is calculable in perturbative QCD, just like the renormalization
scale dependence in $\mu$. It turns out this is true only for
large $k_\perp$.

The $\zeta$-evolution equation for ${\cal Q}(x,k_\perp, \mu,
x\zeta)$ has been derived by Collins and Soper in the large
$\zeta$ limit \cite{ColSop81}. Normally, we use ${\cal Q}(x,
k_\perp)$ in the small $k_\perp$ region. Let us extend this
dependence to large $k_\perp$ and introduce the Fourier (or
impact-parameter) representation,
\begin{equation}
    {\cal Q}(x,b_\perp,\mu,x\zeta) = \int d^2k_\perp
    e^{i\vec{b}_\perp\cdot\vec{k}_\perp}  {\cal Q}(x, k_\perp, \mu, x\zeta)
    \ .
\end{equation}
The above integral should be convergent for non-zero $b_\perp$.
[There are UV divergences when $\vec{b}=0$ which we will not
consider here.] The Collins-Soper evolution equation is
\begin{equation}
    \zeta\frac{\partial}{\partial \zeta}{\cal Q}(x,b,\mu,x\zeta)
      = \big(K(\mu, b)+G(\mu, x\zeta)\big){\cal Q}(x,b,\mu,x\zeta) \ ,
\label{zetarg}
\end{equation}
where $K$ depends on the UV renormalization scale $\mu$ and
infrared impact parameter $b$, and is non-perturbative when $b$ is
large; $G$ is perturbative because $\mu$ and $\zeta$ are hard; and
both are free of gluon and quark mass singularity. The sum $K+G$
is independent of UV scale $\mu$ and hence,
\begin{equation}
         \mu\frac{d}{d\mu} K  = - \gamma_K = -\mu\frac{d}{d\mu}G
         ,
\label{reno}
\end{equation}
where $\gamma_K$ is the cusp anomalous dimension \cite{KorRad92}
and is a series in $\alpha_s$ free of infrared singularities. The
derivation of the above equation in Feynman gauge has been given
in Ref. \cite{Col89,Li97}. In the above equation, any power
correction of $(\Lambda_{\rm QCD}/\zeta)^n$ has been ignored and
hence it is true only when $\zeta\gg \Lambda_{\rm QCD}$.

According to the result in the previous section, $G$ gets a
contribution from diagram 1c only, whereas $K$ gets a contribution
from 2b. The sum is
\begin{equation}
  K(b,\mu) + G(x\zeta,\mu)  = -
  \frac{\alpha_sC_F}{\pi}\ln\frac{x^2\zeta^2b^2e^{2\gamma_E-1}}{4}\ ,
\end{equation}
which is valid when $b^2$ is small and where $\gamma_E$ is the
Euler constant. The one-loop anomalous dimension is then,
\begin{equation}
   \gamma_K = \frac{\alpha_s}{\pi}2C_F \ ,
\end{equation}
which is well known. Using the above renormalization group
equation (\ref{reno}), one can sum over large logarithms
$\ln{\zeta^2b^2}$ in $K+G$ when $b$ is small (otherwise $K$ is
non-perturbative). Substituting the result into Eq.
(\ref{zetarg}), one finds a re-summed double-leading logarithms in
$\zeta b$ (see Sec.V).

\subsection{Factorization of Soft Gluons in
the TMD Parton Distribution}

From the viewpoint of QCD factorization, parton distributions are
introduced to absorb collinear divergences when the quark masses
are zero. From the one-loop result, it is seen that the TMD
distribution contains both collinear and soft contributions, the
latter is manifest through the gluon mass dependence. In this
subsection, we will attempt to isolate and subtract the soft
contribution from the above definition of TMDPD.

Let us first consider the self-energy diagram in Fig. 1. The
diagram 1a contains the soft-gluon contribution which is obtained
by making the soft approximation to the quark propagator and quark
gluon coupling. Briefly, the soft approximation corresponds to
neglecting the soft-gluon momentum in the numerator and the gluon
momentum squared in the denominator; more discussion on the soft
approximation is provided in Sec. IV. In the soft region where all
of the components of $\ell^\mu$ are small, the self-energy becomes
\begin{equation}
   \int \frac{d^4\ell}{(4\pi)^2} \frac{1}{(\ell\cdot p+i\epsilon)^2}
   \frac{1}{(\ell^2-\lambda^2+i\epsilon)} \ .
\end{equation}
We can factorize the above contribution out of the parton
distribution by subtracting it from the one-loop result of $Z_F$.
To make the subtraction mathematically convenient, we extend the
soft approximation to the whole integration region of $\ell$ and
use DR and $\overline{\rm MS}$ scheme to get rid of the UV
contribution. [There is no contribution from the collinear region
because the integral is convergent in the massless quark limit.]
The result is a contribution similar to the self-energy of an
eikonal line.

We can do the similar subtraction for diagram 1b, by forming a
soft approximation for the gluon interacting with the quark line
\begin{equation}
-ig^2\int \frac{d^4\ell}{(2\pi)^4}
   \frac{2p\cdot v}{
(v\cdot\ell +
i\epsilon)(p\cdot\ell+i\epsilon)(\ell^2-\lambda^2+i\epsilon)} \ ,
\label{soft1}
\end{equation}
with $\ell^\mu$ restricted to the soft region. This time, however,
the situation is more complicated. If one extends the integration
$\ell^\mu$ to all regions, there is also a collinear contribution
coming from virtual gluons with momentum parallel to $p^\mu$, as
signified by the divergence of the zero quark mass. In other
words, the simplified approach of subtracting away the whole
integral will also take away a part of the collinear contribution.

One may get around this by excluding the collinear gluon
contribution with small $\ell^-$. This can be achieved by
introducing a four-vector $\tilde v^\mu$ which has a large $\tilde
v^+$ but relatively small $\tilde v^-$, and approximating the
above soft contribution with the following integral,
\begin{equation}
-ig^2\int \frac{d^4\ell}{(2\pi)^4}
   \frac{2\tilde v\cdot v}{
(v\cdot\ell + i\epsilon)(\tilde
v\cdot\ell+i\epsilon)(\ell^2-\lambda^2+i\epsilon)} \ ,
\end{equation}
where we have replaced $p$ by $\tilde v$. The above soft
contribution includes soft gluons with $\ell^+/\ell^-$ limited by
$v^+/v^-$ and $\ell^-/\ell^+$ by $\tilde v^-/\tilde v^+$.

Subtracting the above contribution from the diagram 1b, the
remainder has a soft divergence in the gluon mass. This indicates
that the soft and collinear divergences cannot be completely
separated, as there are regions of loop momentum where collinear
and soft divergences overlap. Therefore, one could in principle
{\it define} Eq. (\ref{soft1}) with un-restricted
$\ell$-integration as the ``soft contribution". With this
approach, the subtracted 1b has no soft divergence. However, as we
have mentioned before, the soft contribution is then not entirely
soft. Two different approaches may be considered as two different
subtraction schemes. Here, we use the first one.

If we follow the above procedure, one finds the complete soft
contribution in terms of the matrix element of Wilson lines,
\begin{equation}
   S(\vec{b}_\perp, \mu^2, \rho) =\frac{1}{N_c} \langle 0|
   {\cal L}^\dagger_{\tilde vil}( \vec{b}_\perp, -\infty)
   {\cal L}^\dagger_{vlj}(\infty;\vec{b}_\perp)
   {\cal L}_{vjk}(\infty;0)
    {\cal L}_{\tilde v ki}(0;-\infty) |0\rangle\ ,
\label{soft}
\end{equation}
where $\rho = \sqrt{v^-\tilde v^+/v^+\tilde v^-}$. we have made
the color indices explicit ($i, j, k, l=1,2,3$).
 The subtracted parton distribution can be defined as
\begin{eqnarray}
  q(x, k_\perp, \mu, x\zeta, \rho)
        = \frac{1}{2}\int
        \frac{d\xi^-}{2\pi}e^{-ix\xi^-P^+}\int
        \frac{d^2b_\perp}{(2\pi)^2} e^{i\vec{b}_\perp\cdot
        \vec{k}_\perp}
        \frac{ \langle P|\overline{\mit \Psi}_v(\xi^-,0,\vec{b}_\perp) \gamma^+
        {\mit \Psi}_v(0)|P\rangle}{S(\vec{b}_\perp, \mu^2, \rho)}  \
        .
\end{eqnarray}
This definition differs from that of Collins \cite{Col03} in that
the soft contribution here has no dependence in $\xi^-$. Moreover,
from our one-loop calculation, it is not clear that the above
distribution has a well-defined limit when $\zeta\rightarrow
\infty$ or $v^\mu$ becomes light-like, $v^2=0$, as claimed in
\cite{Col03}.

Let us calculate the one-loop soft subtraction, shown in Fig. 3.
First, the diagrams with self-energy on all four of the Wilson
lines,
\begin{equation}
     \Delta_{\rm soft} q(x, k_\perp)|_{\rm diag.3a} = -\delta(x-1)\delta^2(\vec{k}_\perp)
     2(Z_W-1)\ ,
\end{equation}
half of which cancels the self-energy of the gauge link in Eq.
(\ref{fig1}).
\begin{figure}
\begin{center} \begin{picture}(400,90)(0,0)
\SetWidth{1.0}

\SetOffset(0,15)
\Line(0,0)(0,30)\Line(3,0)(3,30)\Line(0,30)(20,60)\Line(3,30)(23,60)
\Line(65,30)(65,0)\Line(62,30)(62,0)\Line(65,30)(45,60)\Line(62,30)(42,60)
\GlueArc(12,45)(8,235,418){1.5}{5}
\DashLine(32,65)(32,-5){5}\Text(83,30)[c]{+}\Text(32,-15)[c]{(a)}

\SetOffset(100,15)
\Line(0,30)(0,0)\Line(3,30)(3,0)\Line(0,30)(20,60)\Line(3,30)(23,60)
\Line(65,30)(65,0)\Line(62,30)(62,0)\Line(65,30)(45,60)\Line(62,30)(42,60)
\GlueArc(5,30)(14,255,428){2}{8}
\DashLine(32,65)(32,-5){5}\Text(32,-15)[c]{(b)}\Text(83,30)[c]{+}

\SetOffset(200,15)
\Line(0,30)(0,0)\Line(3,30)(3,0)\Line(0,30)(20,60)\Line(3,30)(23,60)
\Line(65,30)(65,0)\Line(62,30)(62,0)\Line(65,30)(45,60)\Line(62,30)(42,60)
\Gluon(0,20)(65,20){-2}{10}
\DashLine(32,65)(32,-5){5}\Text(32,-15)[c]{(c)}\Text(83,30)[c]{+}

\SetOffset(300,15)
\Line(0,30)(0,0)\Line(3,30)(3,0)\Line(0,30)(20,60)\Line(3,30)(23,60)
\Line(65,30)(65,0)\Line(62,30)(62,0)\Line(65,30)(45,60)\Line(62,30)(42,60)
\Gluon(0,15)(54,45){-2}{9}
\DashLine(32,65)(32,-5){5}\Text(32,-15)[c]{(d)}

\end{picture}
\end{center}
\caption{\it Soft-gluon contribution to the TMD parton
distribution at one-loop order. The double lines represent eikonal
line.}
\end{figure}
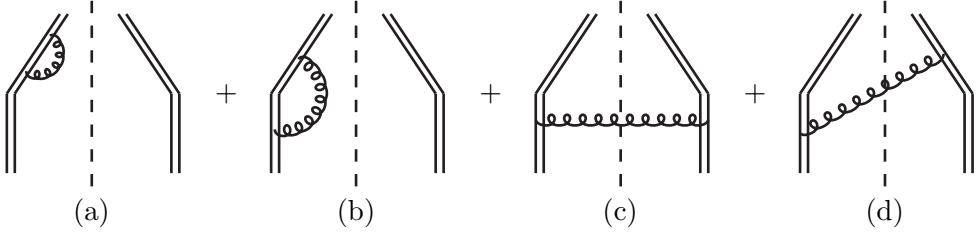

The vertex correction of the gauge links, for which there is a
factor of 2 to account for two virtual vertices, is
\begin{eqnarray}
    \Delta_{\rm soft}q(x, k_\perp)|_{\rm diag.3b}&=& -\delta(x-1)\delta^2(\vec{k}_\perp)2\int \frac{1}{v\cdot
l}\frac{1}{\tilde v\cdot
     l}\frac{1}{l^2-\lambda^2} \nonumber \\
     &=& \delta(x-1)\delta^2(\vec{k}_\perp)\frac{\alpha_s}{2\pi}
     C_F
       \ln\frac{4(v\cdot \tilde v)^2}{v^2\tilde v^2} \cdot
       \ln\frac{\mu^2}{\lambda^2} \ ,
\end{eqnarray}
where the coefficient is just the cusp anomalous dimension
\cite{KorRad92}.

Now we consider the soft contribution from the real emission.
Again there are two types of diagrams. The first type is the one
with two gluons emitted from $v$ or $\tilde v$. The contribution
to the parton distribution is
\begin{equation}
 \Delta_{\rm soft} q (x, k_\perp)|_{\rm diag.3c} =
 \delta(x-1)\frac{\alpha_s}{\pi^2}C_F
 \frac{1}{k_\perp^2+\lambda^2}\ ,
\end{equation}
which, when integrated over $k_\perp$, cancels out the self-energy
contribution. The second type is the interference from the real
gluon emission of the $v$ and $\tilde v$ lines. Plugging in the
soft factor,
\begin{equation}
    \Delta_{\rm soft} q (x, k_\perp)|_{\rm diag.3d}= - \delta(x-1)\frac{\alpha_s}{2\pi^2}
    C_F \ln\frac{4(v\cdot \tilde v)^2}{v^2\tilde v^2}
     \frac{1}{k_\perp^2+\lambda^2} \ .
\end{equation}
When integrated over $k_\perp$, it cancels out the
vertex corrections.

\subsection{Soft Contribution}

According to the definition of the soft contribution in Eq.
(\ref{soft}), the one-loop result for the soft factor can be
constructed from the above soft subtraction contributions,
\begin{eqnarray}
    S(k_\perp, \mu, \rho) &=& \delta^2(k_\perp) - \frac{\sum \Delta_{\rm soft} q(x,
    k_\perp)}{\delta(x-1)} \nonumber \\
    &=&  \delta^2(k_\perp) + \frac{\alpha_s}{2\pi^2}
    C_F \left[\ln\frac{4(v\cdot \tilde v)^2}{v^2\tilde
    v^2}-2\right]\cdot \left[
     \frac{1}{k_\perp^2+\lambda^2} - \pi \delta^2(k_\perp)
     \ln \frac{\mu^2}{\lambda^2}\right]\ .
\end{eqnarray}
When Fourier transformed to $b$-space, it becomes,
\begin{equation}
    S(b, \mu^2, \rho) = 1 + \frac{\alpha_sC_F}{2\pi}(2-\ln\rho^2)
       \ln\left(\frac{\mu^2b^2}{4}e^{2\gamma_E}\right)  \ .
\end{equation}
This, however, cannot be used for $b=0$, for which the integration
over $k_\perp$ must be regularized so that $S(\vec{b}=0)=1$.

\subsection{Integrating over Transverse Momentum in TMDPD}

One would expect that after integrating over transverse momentum,
a TMD parton distribution reduces to the usual Feynman parton
distribution. This, in fact, is not straightforward.

First of all, we have chosen DR and $\overline{\rm MS}$ to
regulate ultraviolet divergences. In the TMDPD, the
$\epsilon\rightarrow 0$ limit and $\overline{\rm MS}$ subtraction
have already been performed as it represents an observable in
4-dimension. On the other hand, a Feynman parton distribution is
obtained first by integrating the transverse-momentum and then
performing UV subtraction. Since the integral is a divergent one,
the procedure of integration and subtraction is not
interchangeable. To get around this, one may use a momentum
cut-off to regularize the UV divergence \cite{BroLep80}; but this
is hard to implement beyond one-loop without destroying the gauge
symmetry. [A consistent regularization might be a discrete
space-time lattice, however, in practice this is hard to implement
in Minkowski space.]

Even when there is a consistent cut-off regulator, one may still
have a problem with the light-cone singularities. While the gauge
link in Feynman distributions is along the light-cone, we have
chosen an off-light-cone gauge link to regulate these divergences.
Therefore, one cannot expect that after integrating over a TMDPD
with a non-light-like gauge the Feynman distribution recovers.

What happens if one takes the light-cone limit of the
non-light-like gauge link after integrating over the transverse
momentum in a TMDPD? The standard integrated parton distribution
still does not emerge if dimensional regularization is used. If
one chooses the non-light-like gauge from the beginning, the gluon
propagator has a term proportional to $v^2$ (if $v$ is the
direction of the gauge link). This term contributes to the TMDPD
in loop calculations. When the light-cone limit, $v^2\rightarrow
0$, is taken, these contributions would vanish if there were no
axial-gauge singularities at $k\cdot v=0$. In practice, however,
the singularities are present and the limit $v^2\rightarrow 0$
does not reproduce the result obtained with $v^2=0$ set in the
beginning in dimensional regularization.

To summarize, it is nontrivial to recover a Feynman parton
distribution by integrating over the transverse-momentum in a
TMDPD. One could cut off the integral by hand, but a UV
regularization scheme must be used which implements the same
cut-off in the loop integrals. This is difficult to construct
beyond one loop. For the same reason, the light-cone limit of the
vector $v^\mu$ is not analytical.

\subsection{TMD Fragmentation Function}

The transverse-momentum-dependent fragmentation function (TMDFF)
has a similar definition as the TMD parton distribution. Many of
the results discussed in the previous subsections can be
immediately translated into those for TMDFF. Here, we sketch the
main results briefly.

Use $\hat q_h(z,k_\perp, \mu, \zeta)$ as a notation for a
subtracted quark fragmentation function into the hadron $h$,
\begin{eqnarray}
  \hat q_h(z, P_{h\perp},\mu,\hat \zeta/z, \rho) &=&\frac{1}{2z}
  \int \frac{d\xi^-}{2\pi}\frac{d^2 \vec{b}}{(2\pi)^2}
  e^{-i(k^+\xi^--\vec{k}_\perp\cdot\vec{b}_\perp)}
 \label{ffdef} \\
  && \times \sum_X \frac{1}{3}\sum_a\langle 0|{\cal L}_{\tilde v}(-\infty;0)\psi_{\beta a}(0)
   |P_hX\rangle \gamma^+_{\alpha\beta}
 \nonumber \\ && \times \langle P_hX|(\overline{\psi}_{\alpha a}(\xi^-,\vec{b}) {\cal
L}_{\tilde v}^\dagger(\xi^-,\vec{b};-\infty)|0\rangle/S(b_\perp,
\mu, \rho) \nonumber \ ,
\end{eqnarray}
where $\tilde v$ is mainly along the light-cone direction
conjugating to $P_h$; $k^+=P^+_h/z$ and $k_\perp =
-\vec{P}_{h\perp}/z$; and $a$ is a color index. The variable $\hat
\zeta$ is defined as
\begin{equation}
         \hat \zeta^2 = 4(P_h\cdot \tilde v)^2/\tilde v^2 \ .
\end{equation}
For a quark fragmenting into a quark, the leading order result is
normalized to $\hat q(z, P_\perp)= \delta(z-1)\delta^2(P_\perp)$.

It is not difficult to see that the quark fragmentation function
in a quark can be obtained by a simple substitution of the
corresponding quark distribution in a quark,
\begin{eqnarray}
    \hat q_h(z, P_\perp, \mu^2, \hat \zeta/z, \rho)
     &=& \frac{1}{z} q\left(\frac{1}{z},\frac{P_\perp}{z}, \mu^2,
     \hat \zeta/z, \rho\right) \nonumber \\
     &=& q(z, P_\perp, \mu^2,z\hat \zeta, \rho) \ ,
\end{eqnarray}
where the second equality holds at one-loop order. Moreover, $
\hat q_h$ satisfies the same Collins-Soper equation in $\hat
\zeta$ evolution as that of $q$ in $\zeta$ evolution.

\section{One-Loop Factorization}

In this section, we show that the factorization formula, Eq. (1),
is valid at one-loop order. To accomplish this, semi-inclusive DIS
on a single quark target is studied. The result can be easily
translated into that for a non-perturbative hadronic target.

In the first subsection, we establish notation and normalization
for the tree level result. In the second subsection, we state and
explain the content of the factorization theorem. In the following
two subsections, we will verify its correctness for a single quark
target on the diagram-by-diagram basis: first for the virtual
corrections, and then for the real corrections.

\subsection{Notation and Tree Normalization}

We choose a coordinate system for semi-inclusive DIS in which the
nucleon is travelling along the $z$-direction. Introduce the
light-cone vectors $(p^0,p^x,p^y,p^z)=\Lambda (1,0,0,1)$,
$(n^0,n^x, n^y,n^z)= (1,0,0,-1)/2\Lambda$, and $p\cdot n=1$, where
$\Lambda$ is an arbitrary parameter. The initial nucleon momentum
$P$ can be written as,
\begin{equation}
      P^\mu = p^\mu + (M^2/2) n^\mu \ ,
\end{equation}
where $M$ is the nucleon mass. The photon momentum is
$q=\ell-\ell'$,where $\ell$ and $\ell'$ are the initial and final
lepton momenta, respectively. We choose the photon momentum in the
negative-$z$ direction,
\begin{equation}
      q^\mu = -\xi p^\mu + \frac{Q^2}{2\xi} n^\mu \ ,
\end{equation}
where $\xi\sim x_B=Q^2/2(P\cdot q)$ when $M^2/Q^2$ is neglected
and $Q^2=-q^2$. The so-called {\it hadron frame} is obtained by
making a particular choice of $\Lambda$ \cite{MenOlnSop92}.

One has the option of either fixing the lepton plane as the $xz$
plane or the hadron plane as the $xz$ plane. In either case, there
is an azimuthal angle between the two planes, and for the
simplicity of our discussion, we integrate out this angle. The
detected hadron has a momentum $P_h$ mainly along the $n^\mu$
(negative $z$) direction with $z_h$ fraction of the photon
momentum component in the same direction, and with transverse
momentum $P_{h\perp}$ which is invariant under the boost along the
$z$-direction. As indicated earlier, $P_{h\perp}$ is considered to
be soft (on the order of $\Lambda_{\rm QCD}$). If
$P_{h\perp}\gg\Lambda_{\rm QCD}$, a different factorization
formula exists in which only the integrated parton distributions
and fragmentation functions enter.

The semi-inclusive DIS cross section under the one-photon exchange
is
\begin{equation}
    \frac{d\sigma}{dx_Bdydz_hd^2P_{h\perp}}
      = \frac{2\pi\alpha^2_{\rm em}}{Q^4}
      y\ell_{\mu\nu}W^{\mu\nu}(P, q, P_h) \ ,
\end{equation}
where the unpolarized lepton tensor is
\begin{eqnarray}
\ell^{\mu\nu} &=& 2(\ell^\mu{\ell'}^\nu + \ell^\mu{\ell'}^\nu -
g^{\mu\nu}Q^2/2) \nonumber \\
&=& (Q^2/y^2)(1-y+y^2/2)(-2g^{\mu\nu}_\perp) + ... \ ,
\end{eqnarray}
where $y$ is the fraction of the lepton energy loss, $1-E'/E$.
Since we are going to integrate over the azimuthal angle $\phi$,
only the structure $g^{\mu\nu}_\perp = g^{\mu\nu}-p^\mu n^\nu -
p^\nu n^\mu$ will survive.

The hadron tensor has the following expression in QCD,
\begin{equation}
   W^{\mu\nu}(P, q, P_h) =\frac{1}{4z_h}\sum_X
   \int \frac{d^4\xi}{(2\pi)^4} e^{iq\cdot \xi}
   \langle P|J_\mu(\xi)|XP_h\rangle \langle XP_h|J_\nu(0)|P\rangle
   \ ,
\end{equation}
where $J^\mu$ is the electromagnetic current of the quarks, $X$
represents all other final-state hadrons other than the observed
particle $h$. The variable $z_h$ can be defined as $P\cdot
P_h/P\cdot q$ or $P_h^-/q^-$.

A simple calculation on the single-quark target yields that,
\begin{equation}
    W^{\mu\nu} = -\frac{1}{2}g^{\mu\nu}_\perp
    \delta(x_B-1)\delta(z_h-1)\delta^2(\vec{P}_{h\perp}) + ...  \
    .
\end{equation}
In this case, it is known
\begin{eqnarray}
      q^{(0)}(x_B, k_\perp) &=& \delta(x_B-1)\delta^2(\vec{k}_\perp) \ ,
      \nonumber\\
      \hat q^{(0)}(z_h, p_{\perp}) &=&
      \delta(z_h-1)\delta^2(\vec{p}_{\perp})\ .
\end{eqnarray}
It is easy to translate the above into a result for a physical
hadron,
\begin{equation}
W^{\mu\nu} = -\frac{1}{2}g^{\mu\nu}_\perp \int d^2\vec{k}_\perp
q(x_B, k_\perp) \int d^2\vec{p}_{\perp} \hat q(z_h, p_{\perp})
\delta^2(z_h\vec{k}_\perp+\vec{p}_\perp -  \vec{P}_{h\perp}) \ .
\end{equation}
Therefore the cross section is,
\begin{eqnarray}
    \frac{d\sigma}{dx_Bdydz_hd^2P_{h\perp}}
      &=& \frac{4\pi\alpha^2_{\rm em}s}{Q^4}(1-y+y^2/2)
      x_B\sum_q e_q^2 \nonumber \\
     & \times & \int
      d^2\vec{k}_\perp
q(x_B, k_\perp) \int d^2\vec{p}_{\perp} \hat q_h(z_h, p_{\perp})
\delta^2(z_h\vec{k}_\perp+\vec{p}_\perp -  \vec{P}_{h\perp}) \ ,
\end{eqnarray}
where $s =(P+\ell)^2$, and we have kept only the
$\phi$-independent term. This result is known in the literature
\cite{TanMul96}.

\subsection{General Form of Factorization}

In the following discussion, we are interested in the leading
structure $F(x_B, z_h,P_{h\perp}, Q^2)$ only,
\begin{equation} W^{\mu\nu} =
-\frac{1}{2}g^{\mu\nu}_\perp F(x_B,z_h,P_{h\perp}, Q^2) + ...\ .
\end{equation}
The other structures factorize in a similar way. The form of the
factorization theorem we want to show is
\begin{eqnarray}
 F(x_B,z_h,P_{h\perp},Q^2)&=&\sum_{q=u,d,s,...} e_q^2\int d^2\vec{k}_\perp d^2\vec{p}_{\perp}
      d^2\vec{\ell}_\perp
   \nonumber \\
   && \times  q\left({x_B}, k_\perp,\mu^2,x_B\zeta, \rho\right)
    \hat q_h\left({z_h}, p_{\perp},\mu^2,\hat\zeta/z_h, \rho\right)
    S(\vec{\ell}_\perp,\mu^2,\rho) \nonumber \\
&& \times H\left(Q^2,\mu^2,\rho\right)
\delta^2(z_h\vec{k}_\perp+\vec{p}_\perp +\vec{\ell}_\perp-
\vec{P}_{h\perp}) \ ,
\end{eqnarray}
where in a special system of coordinates: $\zeta^2 =
(Q^2/x_B^2)\rho$ and $\hat \zeta^2 = (Q^2z_h^2)\rho$ and
$\rho=\sqrt{v^-\tilde v^+/v^+\tilde v^-}$ is a gluon rapidity
cut-off parameter. The above result is accurate up to powers in
$(P_{h\perp}^2/Q^2)^n$ for soft $P_{h\perp}\sim \Lambda_{\rm
QCD}$. There is no direct contribution from the gluon distribution
in this kinematic region. There is no convolution involving the
longitudinal momentum fractions, $x_B$ and $z_h$, typical in other
hard processes. The transverse-momentum integrals show that the
hadron transverse-momentum can be generated from the initial state
parton, final state fragmentation, and the soft gluon radiation.

The renormalization and collinear factorization scale $\mu$
cancels among the four factors, as the structure function should
be $\mu$-independent. As in the inclusive case, one can choose
$\mu^2=Q^2$ to eliminate the large logarithms in the hard factor.
The soft-factorization parameter $\rho$ depends on the directions
of the Wilson lines and must also be cancelled among all factors.
The Collins-Soper equation allows studying the double logarithmic
dependence of $Q^2$ in the distribution and fragmentation
functions.

In principle, for $P_{h\perp}\sim Q$, the above factorization
formula breaks down because of the power corrections, and because
the transverse-momentum $P_{h\perp}$ is now mainly generated from
multi-jets production. However, it is convenient to extrapolate
the above factorization to all $P_{h\perp}$, and introduce the
impact-parameter space representation,
\begin{eqnarray}
   F(x_B,z_h,b, Q^2) &=& \sum_{q=u,d,s,...} e_q^2
      q\left(x_B, z_hb, \mu^2, x_B\zeta, \rho\right)\hat q\left(z_h,b,\mu^2,
   \hat \zeta/z_h, \rho\right) \nonumber \\
   && \times S(b,\mu^2, \rho) H\left(Q^2,\mu^2,\rho \right)\ .
\end{eqnarray}
The convolution in the transverse momentum becomes a product of
Fourier factors.

At tree level, $S^{(0)}(b,\mu^2, \rho)=1$, and $H^{(0)}=1$. Let us
show that at one-loop, the above factorization is still valid, and
calculate the correction to the hard part at this order.

\subsection{Virtual Corrections}

We first consider factorization of one-loop virtual corrections to
the tree process. The factorization actually holds diagram by
diagram, and therefore we will study the momentum flow in
individual diagrams and extract the corresponding hard factor.

Three diagrams shown in Fig. 4 correspond to the initial and final
state wave function renormalization and vertex corrections.

\begin{figure}
\begin{center} \begin{picture}(330,90)(0,0)
\SetWidth{1.0}

\SetOffset(0,15)
\Photon(-10,70)(0,50){2}{5}\Photon(70,70)(60,50){2}{5}
\ArrowLine(0,0)(0,15)\ArrowLine(0,35)(0,50)\Line(0,15)(0,35)
\GlueArc(0,25)(10,-90,90){2.0}{5}\Line(0,50)(60,50)
\ArrowLine(60,50)(60,0)\DashLine(35,65)(35,-5){5}\Text(85,25)[c]{+}
\Text(30,-15)[c]{(a)}

\SetOffset(110,15)
\Photon(-10,70)(0,50){2}{5}\Photon(70,70)(60,50){2}{5}\ArrowLine(0,0)(0,50)
\ArrowLine(0,50)(30,50)\GlueArc(20,50)(10,180,360){2.0}{5}\Line(30,50)(60,50)
\ArrowLine(60,50)(60,0)\DashLine(40,65)(40,-5){5}
\Text(85,25)[c]{+}\Text(30,-15)[c]{(b)}

 \SetOffset(220,15)\Photon(-10,70)(0,50){2}{5}\Photon(70,70)(60,50){2}{5}
\ArrowLine(0,0)(0,20)\ArrowLine(0,20)(0,50)
\ArrowLine(0,50)(30,50)\Gluon(0,20)(30,50){-2.0}{6}\Line(30,50)(60,50)
\ArrowLine(60,50)(60,0)\DashLine(40,65)(40,-5){5}
\Text(30,-15)[c]{(c)}
\end{picture}
\end{center}
\caption{\it One-loop virtual correction to semi-inclusive DIS.}
\end{figure}
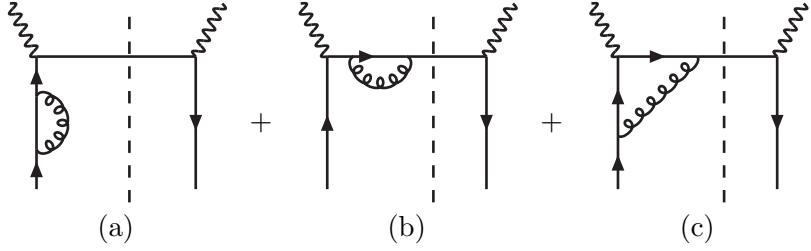

The self-energy correction is straightforward,
\begin{equation}
    F = \delta(x_B-1)\delta(z_h-1)\delta^2(\vec{P}_{h\perp})
     (1+2(Z_F-1))+ ...
\end{equation}
where $Z_F$ is given in Eq. (\ref{zf}). $Z_F$ contains both soft
and collinear contribution in the on-shell scheme. If the
self-energy is associated with the initial state quark, the
collinear part of $Z_F$ is absorbed by the self-energy correction
on the parton distribution, corresponding to diagram 1a subtracted
by diagram 3a. The remaining soft contribution is attributed to
the soft factor shown in diagram 3a. There is no contribution from
the fragmentation function, nor is there a contribution from the
hard part. A similar argument shows the self-energy correction to
the final-state quark can be absorbed by the fragmentation
function and the soft factor, yielding no contribution to the hard
part.

The vertex correction produces exactly the same expression for $F$
as the above, except $Z_F$ is replaced by,
\begin{equation}
    \hat Z_V = 1
    -\frac{\alpha_s}{4\pi}C_F\left(\ln\frac{Q^2}{\mu^2}
   + \ln^2\frac{Q^2}{m^2} +
    2\ln\frac{m^2}{\lambda^2}\ln\frac{Q^2}{m^2}
    -4\ln\frac{Q^2}{m^2}-\frac{\pi^2}{3}\right)
\end{equation}
The UV divergence in the above expression cancels that in $Z_F$
because the sum of one-loop virtual corrections has no UV
divergence. The above result contains a quark distribution part
shown in Fig. 1c subtracted by Fig. 3b, and a fragmentation
function with a similar structure, and a soft contribution in Fig.
3b. Subtracting all of the above from $\hat Z_V-1$, we find a
left-over hard contribution,
\begin{equation}
    H^{(1)}(Q^2,\mu^2,\rho) = \frac{\alpha_s}{2\pi}C_F\left[\left(1+\ln\rho^2\right)
     \ln\frac{Q^2}{\mu^2}
     - \ln\rho^2 + \frac{1}{4}\ln^2\rho^2 + \pi^2-4\right] \ ,
\end{equation}
where we have chosen a coordinate system in which $x_B\zeta =
\hat\zeta/z_h$ and therefore the dependence on the
quasi-light-like vectors $v$ and $\tilde v$ is simply through a
combination, $\rho = \sqrt{v^-\tilde v^+/v^+\tilde v^-}$. The
dependence on $Q^2$ is of the form of single logarithms and can be
controlled by a renormalization group equation because it contains
no additional scale other than $\mu$.

\subsection{Real Corrections}

The one-loop real corrections are shown in Fig. 4. Since we are
interested in the topology of the final state in which the struck
quark carries the dominant part of the energy-momentum of the
current region, the emitted gluons are considered to be either
soft or collinear. Therefore, there is no contribution to the hard
scattering kernel from any of these diagrams. Our job is to show
that these diagrams can be properly taken into account by the
known one-loop parton distribution, fragmentation function, and
the soft factor.

Let us start with the ladder diagram shown in Fig. 5a. The
soft-gluon radiation generates a transverse-momentum for the
struck quark. There is no contribution from the fragmentation
function because the contribution from the final state with a
gluon in the $n^\mu$ direction and a soft quark is power
suppressed. Therefore, the diagram must be factorizable into the
parton distribution in Fig. 2a subtracted off Fig.3c, and the soft
factor in Fig. 3c. A simple calculation of the diagram yields,
\begin{eqnarray}
    F &=& \frac{\alpha_s}{2\pi^2}C_F \delta(z_h-1)
  (1-x_B) \left[\frac{1}{P_{h\perp}^2 + x_B\lambda^2 +
  (1-x_B)^2m^2} \right. \nonumber \\
   && ~~~~~ \left.- \frac{2xm^2}{(P_{h\perp}^2+ x_B\lambda^2 +
   (1-x_B)^2m^2)^2}\right]\ .
\end{eqnarray}
Indeed, the above expressions can easily be reproduced by the
factorization formula with a one-loop result for $q^{(1)}$ and
$S^{(1)}$ and tree level $\hat q^{(0)}$ and $H^{(0)}$.

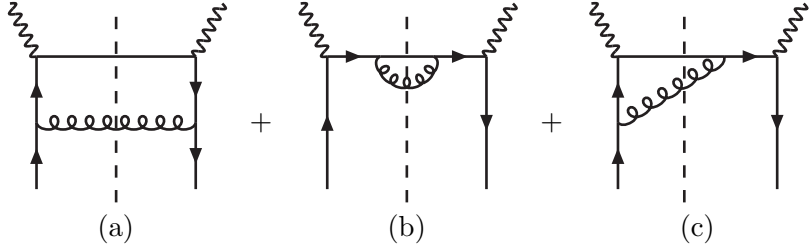
\begin{figure}
\begin{center} \begin{picture}(330,90)(0,0)
\SetWidth{1.0}

\SetOffset(0,15)
\Photon(-10,70)(0,50){2}{5}\Photon(70,70)(60,50){2}{5}
\ArrowLine(0,0)(0,25)\ArrowLine(0,25)(0,50)
\Gluon(0,25)(60,25){-2.5}{8}\Line(0,50)(60,50)
\ArrowLine(60,50)(60,25)\ArrowLine(60,25)(60,0)\DashLine(30,65)(30,-5){5}
\Text(85,25)[c]{+} \Text(30,-15)[c]{(a)}

\SetOffset(110,15)
\Photon(-10,70)(0,50){2}{5}\Photon(70,70)(60,50){2}{5}\ArrowLine(0,0)(0,50)
\ArrowLine(0,50)(20,50)\ArrowLine(40,50)(60,50)\Line(20,50)(40,50)
\GlueArc(30,50)(10,180,360){2.0}{5}
\ArrowLine(60,50)(60,0)\DashLine(30,65)(30,-5){5}
\Text(85,25)[c]{+}\Text(30,-15)[c]{(b)}

 \SetOffset(220,15)\Photon(-10,70)(0,50){2}{5}\Photon(70,70)(60,50){2}{5}
 \ArrowLine(0,0)(0,25)\ArrowLine(0,25)(0,50)
\Gluon(0,25)(40,50){-2.5}{6}\ArrowLine(40,50)(60,50)\Line(0,50)(40,50)
\ArrowLine(60,50)(60,0)\DashLine(25,65)(25,-5){5}
\Text(30,-15)[c]{(c)}
\end{picture}
\end{center}
\caption{\it One-loop real correction to semi-inclusive DIS.}
\end{figure}

Similarly for Fig. 4b, we have
\begin{eqnarray}
    F &=& \frac{\alpha_s}{2\pi^2}C_F \delta(x_B-1)
  (1-z) \left[\frac{1}{P_{h\perp}^2 + z_h\lambda^2 +
  (1-z_h)^2m^2}\right.
  \nonumber \\
   && ~~~~~
   \left. - \frac{2xm^2}{(P_{h\perp}^2+ z_h\lambda^2 +
(1-z_h)^2m^2)^2}\right] \ ,
\end{eqnarray}
which again can be reproduced by the factorization formula with
the one-loop fragmentation function and the soft factor S, and the
tree-level parton distribution and the hard part.

Finally, let us consider the diagram Fig. 5c and its hermitian
conjugate. In the region where $P_{h\perp}$ is small, we find
three distinct contributions:
\begin{eqnarray}
 F &=&  \frac{\alpha_s
    C_F}{2\pi^2} \delta(z_h-1)\frac{2x_B}{(1-x_B)_+}
    \left[\frac{1}{P_{h\perp}^2 + x_B\lambda^2 + (1-x_B)^2m^2}\right]
    \nonumber \\
&&+ \frac{\alpha_s
    C_F}{2\pi^2} \delta(x_h-1)\frac{2z_h}{(1-z_h)_+}
    \left[\frac{1}{P_{h\perp}^2 + z_h\lambda^2 + (1-z_h)^2m^2}\right]
    \nonumber \\
&& + \frac{\alpha_s
    C_F}{2\pi^2} 2\delta(x_B-1)\delta(z_h-1)\frac{1}{P_{h\perp}^2+
    \lambda^2} \ln \frac{Q^2}{P_{h\perp}^2+
    \lambda^2} \ ,
\end{eqnarray}
where the first term corresponds to a gluon collinear to the
initial quark, the second term a gluon collinear to the final
state quark, and the third term a soft gluon. All these terms are
reproduced by the factorization formula with one-loop parton
distribution, fragmentation function, and the soft factor.

Therefore we conclude that at the one-loop level, the general
factorization formula holds.

\section{Factorization to All-Orders in Perturbation Theory}

In this section, we argue that the factorization formula we stated
in the previous section holds to all orders in perturbative QCD.
To make such arguments, we follow the steps outlined in an
excellent review article by Collins, Sterman, and Soper
\cite{CSS89}. One must consider a general Feynman diagram and
study its leading contributions to the SIDIS cross section. The
contributions from different regions of the internal momentum
integrations are characterized by the reduced diagrams which
correspond to pinched surfaces in the space of integration
variables. The leading reduced diagrams can be determined by
infrared power counting. The remaining steps involve decoupling
the Lorentz and color indices, and using soft approximation to
disentangle momentum integrals in the different parts of the
reduced diagrams. To simplify the derivation, one must use the
(generalized) Ward identities extensively. After decoupling and
replacing the various factors by the parton distribution,
fragmentation function and the soft function, one finally arrives
at the general form of factorization.

Using the fact that the physical observables are independent of
renormalization and soft-collinear factorization scales, large
double and single logarithms involved in the factorization formula
can be summed. The final expression is useful to describe
experimental data when combined with result from perturbative
calculations of a fixed order.

\subsection{Reduced Diagrams, Power Counting, and Leading Regions}

\begin{figure}
\SetWidth{0.7}
\begin{center} \begin{picture}(170,140)(0,0)
\SetOffset(20,10) \SetWidth{0.7}
\Photon(2,70)(-20,90){2}{4}\Photon(138,70)(160,90){2}{4}
\Text(-22,92)[r]{$\gamma^*$}\Text(165,94)[l]{$\gamma^*$}
\GOval(70,10)(60,8)(90){0.7} \COval(70,55)(15,10)(90){Black}{Red}
\SetColor{Red}
\DashLine(40,16)(60,47){2}\DashLine(100,16)(80,47){2}
\DashLine(60,63)(50,100){2}\DashLine(80,63)(90,100){2}
\DashLine(22,66)(55,55){2}
\DashLine(118,66)(85,55){2}\SetColor{Black}

\ArrowLine(10,10)(10,56)\ArrowLine(130,56)(130,10)
\ArrowLine(6,10)(6,56)\ArrowLine(14,10)(14,56)
\ArrowLine(12,75)(40,106)\ArrowLine(100,106)(128,75)
\ArrowLine(15,72)(43,103)\ArrowLine(8,76)(39,110)
\ArrowLine(97,103)(125,72)\ArrowLine(101,110)(132,76)
\ArrowLine(126,56)(126,10)\ArrowLine(134,56)(134,10)
\COval(10,66)(12,10)(90){Black}{Blue}\COval(130,66)(12,10)(90){Black}{Blue}
\GOval(70,10)(64,8)(90){0.3} \GOval(70,106)(35,7)(90){0.3}
\DashLine(70,130)(70,-10){5} \SetWidth{1.8}\ArrowLine(-12,0)(6,10)
\ArrowLine(134,10)(152,0)
\ArrowLine(35,106)(45,125)\ArrowLine(95,125)(105,106)
\Text(70,10)[c]{$J_t$} \Text(70,106)[c]{$J_c$}
\Text(10,66)[c]{$H$} \Text(130,66)[c]{$H$}\Text(70,55)[c]{$S$}
\Text(-14,0)[r]{$P$}\Text(154,0)[l]{$P$}
\Text(50,130)[b]{$P_h$}\Text(90,130)[b]{$P_h$}\end{picture}
\end{center}
\caption{\it A general reduced diagram for semi-inclusive DIS.}
\end{figure}
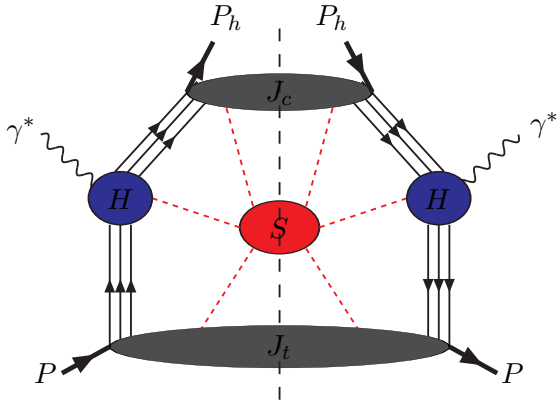

The contribution of an arbitrary (cut) Feynman diagram to the
SIDIS cross section can be classified in term of pinched surfaces
corresponding to the solutions of Landau equations
\cite{Lan59,Ste78,LibSte78}. Coleman and Norton observed that
these pinched surfaces can be pictured in terms of physical
space-time processes (or reduced diagrams) \cite{ColNor65}.
According to the kinematic constraints of SIDIS, it is not
difficult to see that the most general reduced diagrams have the
structure shown in Fig. 6, in which the initial nucleon evolves
into a target fragmentation jet $J_t$ plus a set of collinear
quarks and gluons (solid lines) entering the hard-interaction
vertex with the highly virtual photon $\gamma^*$. A new set of
collinear quarks and gluons (solid lines) emerges from the hard
vertex in a new direction (opposite direction in the collinear
frames), and fragments into the observed hadron and the unobserved
jet $J_c$. Fig. 6 is actually a cut diagram including the
complex-conjugated amplitudes, corresponding to the measured cross
section. Therefore, we will use additional indices $L$ and $R$ to
label jets on the left and right sides of the cut (indicated by
the vertical dashed line), respectively. For example, $J_{cR}$
labels the current jet on the right-hand-side of the cut. In
addition, there is a soft sub-diagram $S$ with soft quark and
gluon lines (shown by dashed lines) connecting the jets and hard
parts.

Let us count the degree of infrared divergence $\omega(G)$ of each
reduced diagrams $G$. It can be constructed from the sum of the
degrees of divergences for the jets and the soft part,
\begin{equation}
    \omega(G) = \omega_{J_{tL}} + \omega_{J_{tR}}
     + \omega_{J_{cL}} + \omega_{J_{cR}} + \omega_S \ .
\end{equation}
The power counting for the soft function is straightforward: If we
use $E^b$ and $E^f$ to denote the number of soft boson (gluon) and
fermion external lines, then it is well known that
\begin{equation}
       \omega_S=E^b+\frac{3}{2} E^f
\end{equation}
from a simple dimensional analysis in coordinate space. Note that
$\omega_S$ includes the propagators of the external lines and the
associated integration measure.

\begin{figure}
\SetWidth{0.7}
\begin{center} \begin{picture}(170,140)(0,0)
\SetOffset(20,10) \SetWidth{0.7}
\Photon(2,70)(-20,90){2}{4}\Photon(138,70)(160,90){2}{4}
\GOval(70,10)(60,8)(90){0.7} \COval(70,55)(15,10)(90){Black}{Red}
\SetColor{Red}
\Gluon(40,16)(60,47){2}{6}\Gluon(100,16)(80,47){-2}{6}
\Gluon(60,63)(50,100){2}{6}\Gluon(80,63)(90,100){-2}{6}
\SetColor{Black}

\Line(10,10)(10,56)\Line(130,10)(130,56)
\Gluon(6,10)(6,56){-2}{6}\Gluon(14,10)(14,56){-2}{6}
\Line(12,75)(40,106)\Line(128,75)(100,106)
\Gluon(13,71)(41,102){-1.5}{7}\Gluon(8,76)(36,107){-1.5}{7}
\Gluon(126,72)(98,103){1.5}{7}\Gluon(132,76)(104,107){1.5}{7}
\Gluon(126,10)(126,56){2}{6}\Gluon(134,10)(134,56){2}{6}
\COval(10,66)(12,10)(90){Black}{Blue}\COval(130,66)(12,10)(90){Black}{Blue}
\GOval(70,10)(64,8)(90){0.3} \GOval(70,106)(35,7)(90){0.3}
\DashLine(70,130)(70,-10){5} \SetWidth{1.8}\ArrowLine(-12,0)(6,10)
\ArrowLine(134,10)(152,0)
\ArrowLine(35,106)(45,125)\ArrowLine(95,125)(105,106)
\Text(70,10)[c]{$J_t$} \Text(70,106)[c]{$J_c$}
\Text(10,66)[c]{$H$} \Text(130,66)[c]{$H$}\Text(70,55)[c]{$S$}
\Text(-14,0)[r]{$P$}\Text(154,0)[l]{$P$}
\Text(55,125)[b]{$P_h$}\Text(85,125)[b]{$P_h$}\end{picture}
\end{center}
\caption{\it The leading region for semi-inclusive DIS.}
\end{figure}
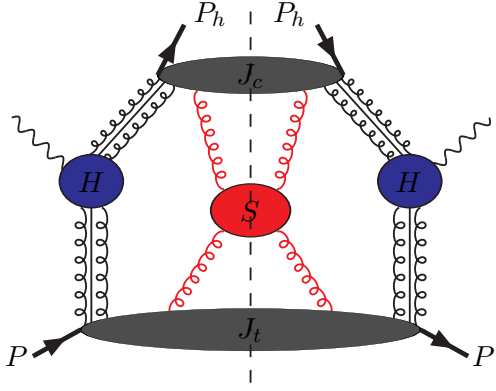

Let us use $p_{J}$ to denote the number of collinear quark or
gluon-with-physical-polarization lines entering the hard part from
jet $J$; use $l_{J}$ to represent the number of collinear gluons
of longitudinal polarization through a similar attachment; use
$E_J^{b,f}$ to denote the number of soft boson or fermion lines
connecting the soft part to the jets; use $E_{HL,R}^{b,f}$ to
label the number of soft bosons or fermions connecting to the left
or right hard part; and finally use $v_J^{(3)}$ to label the
number of three-point vertices in the jet, and $s_J$ the number of
soft gluons with scalar polarization attaching to the jet. Then it
is easy to see that
\begin{eqnarray}
      E^b &=& E^b_{J_{tL}}+E^b_{J_{tR}}+
  E^b_{J_{cL}}+E^b_{J_{cR}} + E^b_{HL}+E^b_{HR}  \ , \nonumber \\
      E^f &=& E^f_{J_{tL}}+E^f_{J_{tR}}+
  E^f_{J_{cL}}+E^f_{J_{cR}} + E^f_{HL}+E^f_{HR} \ .
\end{eqnarray}
The soft power associated with each collinear jet is
\begin{equation}
      \omega_J = 2L_J-N_J + t_J \ ,
\end{equation}
where $L_J$ is the number of loops in the jet (each contributing
two powers), and $N_J$ is the number of internal lines (each
contributing one power), and $t_J$ is the numerator suppression
factor which in Feynman gauge is equal to ${\rm
max}[v_{J}^{(3)}-l_J-s_J, 0]/2$ \cite{Ste78}. The number of loops
can be calculated using
\begin{equation}
      N_J - v_J^{(3)} - v_J^{(4)}  = L_J \ ,
\end{equation}
where $v_J^{(3,4)}$ are the number of three-and four-point
vertices, respectively. The relation between the number of
vertices and lines is,
\begin{equation}
      3v_J^{(3)} + 4v_J^{(4)} + p_J + l_J = 2N_J + E^b_J + E^f_J +
      I_J \ ,
\end{equation}
where $I_J$ is the number of (initial and final) external lines in
the jet. From the above, it is easy to see that
\begin{equation}
  \omega_J = \frac{1}{2}(p_J-s_J - E_J^b - E_J^f - I_J) +
  \frac{1}{2}\left(s_J+l_J-v_J^{(3)}\right)\theta\left(s_J+l_J-v_J^{(3)}\right)
  \ .
\end{equation}
Combining the results from four jets, one finds,
\begin{eqnarray}
  \omega(G)&\ge&
  \frac{1}{2}\left(p_{J_{tL}}+p_{J_{tR}}+
  p_{J_{cL}}+p_{J_{cR}}\right) - \frac{1}{2}(I_t+I_c) \nonumber \\
  &&  + \frac{1}{2}\left(E^b_{J_{tL}}+E^b_{J_{tR}}+
  E^b_{J_{cL}}+E^b_{J_{cR}}\right) \nonumber \\
 &&  - \frac{1}{2}\left(s_{J_{tL}}+s_{J_{tR}}+
  s_{J_{cL}}+s_{J_{cR}}\right) \nonumber \\
    && + E^f_{J_{tL}}+E^f_{J_{tR}}+
  E^f_{J_{cL}}+E^f_{J_{cR}} \nonumber \\
  && + E^b_{HL}+E^b_{HR} + \frac{3}{2}(E^f_{HL}+E^f_{HR})
  \nonumber \\
  && + \frac{1}{2}\left(s_{J_{tL}} + l_{J_{tL}} -
  v_{J_{tL}}^{(3)}\right)\theta \left(s_{J_{tL}} + l_{J_{tL}} -
  v_{J_{tL}}^{(3)}\right) \nonumber \\
    && + \frac{1}{2}\left(s_{J_{tR}} + l_{J_{tR}} -
  v_{J_{tR}}^{(3)}\right)\theta \left(s_{J_{tR}} + l_{J_{tR}} -
  v_{J_{tR}}^{(3)}\right) \nonumber \\
    && + \frac{1}{2}\left(s_{J_{cL}} + l_{J_{cL}} -
  v_{J_{cL}}^{(3)}\right)\theta \left(s_{J_{cL}} + l_{J_{cL}} -
  v_{J_{cL}}^{(3)}\right) \nonumber \\
    && + \frac{1}{2}\left(s_{J_{cR}} + l_{J_{cR}} -
  v_{J_{cR}}^{(3)}\right)\theta \left(s_{J_{cR}} + l_{J_{cR}} -
  v_{J_{cR}}^{(3)}\right) \ .
\end{eqnarray}
From the above, the largest possible degree of infrared divergence
is $0$ if the initial and final state hadrons are replaced by a
perturbative parton $(I_t=I_c=1)$.

According to the above result for $\omega(G)$, leading reduced
diagrams (leading region) must satisfy the following conditions:
\begin{itemize}
\item{No soft fermion lines,} \item{No soft gluon lines attached
to the hard parts,} \item{Soft gluon lines attached to jets must
be longitudinally polarized,} \item{In each jet, one quark line
plus an arbitrary number of longitudinally polarized gluons
attached to the corresponding hard part,}\item{The number of
three-point vertices in a jet must be larger or equal to the
number of soft and longitudinally polarized gluon attachments.}
\end{itemize}
In Fig. 7, we show the leading reduced diagrams satisfying the
above conditions. As indicated already, the collinear gluons are
longitudinally polarized.

\subsection{Factorization of Collinear Gluons}

Let us first factorize the longitudinally-polarized colliner
gluons from the hard parts. This can be done using the approach
discussed in \cite{CSS89}. For definiteness, let us consider the
collinear gluons from the initial state nucleon. Because the
gluons are longitudinally polarized, the gluon gauge potential can
be replaced by
\begin{equation}
    A^\mu = A\cdot n p^\mu \ ,
\end{equation}
where $p^\mu$ is the light-cone momentum to which the initial
nucleon momentum is proportional. The effects of these gluons on
the hard part can be factorized through the Ward identity,
\begin{equation}
  \langle f|T \partial_{\mu_1} A^{\mu_1}(\xi_1)
              \partial_{\mu_2} A^{\mu_2}(\xi_2)
               ... \partial_{\mu_n} A^{\mu_n}(\xi_n)|i\rangle =0 \ ,
\end{equation}
where $|i\rangle$ and $|f\rangle$ are physical states. Applying
this identity repeatedly leads to the conclusion that the
collinear gluons can be viewed as attaching to an eikonal line in
the conjugating light-cone direction $n^\mu$. This result can be
understood in an intuitive way: The longitudinally polarized
gluons cannot resolve the internal dynamics of the hard
scattering. It can, however, be sensitive to the overall flow of
the color charge. The hard interaction is a light-cone dominated
process along the $n^\mu$ direction in the coordinate space. This
is also the direction along which the final state jet is formed.
Thus the collinear gluons mainly scatter with the color-charge
flow in this direction.

The Feynman momentum $x$ of a collinear gluon has a lower limit in
a physical process. For example, the smallest $x$ that a gluon may
have is on the order of $M/Q$. Only in the limit $Q\rightarrow
\infty$, can there be near zero-momentum gluons participating in
the scattering. When a collinear gluon has a small $x$, its
light-cone energy is large, and its contribution to the cross
section can be calculated perturbatively. Therefore, one can
introduce a parameter that separates contributions of the gluons
with different rapidities. The collinear gluons with $x$ larger
than a certain cut-off are included in the parton distributions;
others are included in the hard factor. Of course, the physical
cross section is independent of this parameter. In the inclusive
case, the singular contribution from small-$x$ gluons cancels
between the real and virtual diagrams.

To define a parton distribution with virtual gluons of limited
rapidity, one can introduce a rapidity cut-off. The most
straightforward approach is to implement a lower cut-off in $x$. A
more convenient approach, as we discussed in the one-loop case, is
to introduce a quasi-light-cone vector, $v^\mu$, which is close
but not exactly in the $n^\mu$ direction, and to assume that all
collinear gluons couple to a colored jet moving in this direction.
It can be checked that in this approach only the gluons with
$k^+/k^-> v^+/v^-$ are included in the parton distribution.

The collinear gluons from $J_t$ can be factorized in a similar
way. Here a quasi-light-cone vector $\tilde v$ must be introduced
to limit the small-$x$ gluon contribution to the TMD fragmentation
function: only collinear gluons with $k^-/k^+>\tilde v^-/\tilde
v^+$ are included in the non-perturbative function. The left and
right parts of the cut diagrams can be treated in a symmetric way.

\subsection{Soft Approximation and Soft Factor}

The soft gluons are attached to the target and current jets, and
can be factorized using the Grammer-Yennie approximation
\cite{GraYen73} (or soft approximation). The approximation
consists of two steps. The first step is to neglect any soft
momentum in the numerators of the jet factors. One of the
consequences is that the gluon polarization is effectively along
the conjugating light-cone direction of the jet (longitudinally
polarized). The second step is to neglect $k^2$ compared to $
k\cdot n k\cdot p$ in the jet denominator. This approximation is
not uniformly true in the soft region. In fact, in the so-called
Glauber region, where $k_\perp^2\gg k\cdot nk\cdot p$, the
approximation fails \cite{BBL81}. If, however, the momentum $k^+$
or $k^-$ is not trapped, one can deform the contour integration to
a region where $k\cdot n k\cdot p \gg k_\perp^2$, so that the
approximation can still be applied. It is known that for
semi-inclusive hadron production in $e^+e^-$ annihilation, the
deformation can be easily performed \cite{CSS89}. In inclusive
Drell-Yan, this happens only after summing the final state
interaction diagrams \cite{CSS85}. In the present case, the soft
gluons interact with the current and target jets; all of these
interactions are final state interactions.  Hence, all physical
poles appear in the upper-half plane. As such, the contour
deformation can be done straightforwardly.

After the soft approximation, one can again use the Ward identity
to factorize all the soft gluons from the jets. The physical
effect of a jet can be replaced by a Wilson line along the jet
direction. Again to avoid the light-cone singularity, the Wilson
line can be chosen to be off the light-cone along the $v$ or
$\tilde v$ direction. After factorizing the gluons from the jets,
and summing over all soft contributions, a soft factor emerges:
\begin{equation}
   S(\vec{b}_\perp, \mu^2, \rho) = \frac{1}{N_c} {\rm Tr}
    \langle 0|
   {\cal L}^\dagger_{\tilde v}( \vec{b}_\perp, -\infty)
   {\cal L}^\dagger_{v}(\infty;\vec{b}_\perp)
   {\cal L}_{v}(\infty;0)
    {\cal L}_{\tilde v}(0;-\infty) |0\rangle\ ,
\end{equation}
which appears as a factor in the factorization theorem. Now the
leading region has the form shown in Fig. 8.

\begin{figure}
\SetWidth{0.7} \begin{center} \begin{picture}(170,140)(0,0)
\SetOffset(20,10) \SetWidth{0.7}
\Photon(2,70)(-20,90){2}{4}\Photon(138,70)(160,90){2}{4}
\Text(-22,92)[r]{$\gamma^*$}\Text(165,94)[l]{$\gamma^*$}
\GOval(70,10)(60,8)(90){0.7} \COval(70,55)(11,7)(90){Black}{Red}
\SetColor{Red}
\Gluon(52,39)(65,50){1.5}{3}\Gluon(88,39)(75,50){1.5}{3}
\Gluon(52,71)(65,60){1.5}{3}\Gluon(88,71)(75,60){1.5}{3}
\SetColor{Black} \Line(40,55)(62,25)\Line(43,55)(65,25)
\Line(40,55)(62,85)\Line(43,55)(65,85)
\Line(100,55)(78,25)\Line(97,55)(75,25)
\Line(100,55)(78,85)\Line(97,55)(75,85)
\ArrowLine(10,10)(10,56)\ArrowLine(130,56)(130,10)
\Line(10,56)(25,56)\Line(10,54)(25,54)
\Line(130,56)(115,56)\Line(130,54)(115,54)
\Line(10,75)(33,75)\Line(10,73)(30,73)
\Line(130,75)(107,75)\Line(130,73)(110,73)
\Gluon(25,10)(25,56){-2}{6}\Gluon(18,10)(18,56){-2}{6}
\Gluon(115,10)(115,56){2}{6}\Gluon(123,10)(123,56){2}{6}
\ArrowLine(12,75)(40,106)\ArrowLine(100,106)(128,75)
\Gluon(30,73)(58,106){-1.5}{7}\Gluon(22,73)(50,106){-1.5}{7}
\Gluon(110,73)(82,106){1.5}{7}\Gluon(118,73)(90,106){1.5}{7}
\COval(10,66)(12,10)(90){Black}{Blue}\COval(130,66)(12,10)(90){Black}{Blue}
\GOval(70,10)(64,8)(90){0.7} \GOval(70,106)(35,7)(90){0.7}
\DashLine(70,130)(70,-10){5} \SetWidth{1.8}\ArrowLine(-12,0)(6,10)
\ArrowLine(134,10)(152,0)
\ArrowLine(35,106)(45,125)\ArrowLine(95,125)(105,106)
\Text(70,10)[c]{$J_t$} \Text(70,106)[c]{$J_c$}
\Text(10,66)[c]{$H$} \Text(130,66)[c]{$H$}\Text(70,55)[c]{$S$}
\Text(-14,0)[r]{$P$}\Text(154,0)[l]{$P$}
\Text(50,130)[b]{$P_h$}\Text(90,130)[b]{$P_h$}\end{picture}
\end{center}
\caption{\it The leading region for SIDIS after soft and collinear
factorizations.}
\end{figure}
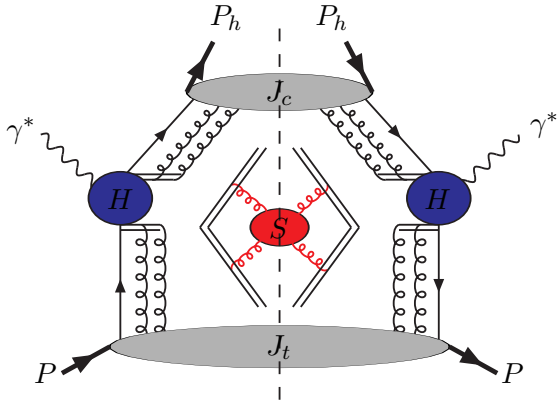

The soft factor is renormalization-scale dependent. The
renormalization group equation is
\begin{equation}
    \mu \frac{\partial S(\vec{b}_\perp, \mu^2, \rho)}{\partial\mu}
       = \gamma_S(\rho) S(\vec{b}_\perp, \mu^2, \rho) \ ,
\end{equation}
where $\gamma_S(\rho)$ is the anomalous dimension of the Wilson
lines in the definition. At one-loop order, one has
\begin{equation}
  \gamma_S =  \frac{\alpha_s}{\pi}
    C_F \left[2-\ln\rho^2\right] + ...
\end{equation}
which is $\rho$ dependent. The anomalous dimension at higher-order
has been studied in Ref. \cite{KorRad92,BotSte89}.

\subsection{Subtracted and Unsubtracted Parton Distributions and Fragmentation Functions}

Let us consider the target jet factor which has been factored from
the hard part, with the soft factor factorized out as well. It is
shown on the left-hand side in Fig. 9. The internal loop momenta
of the gluons are restricted. Their $k^+$ components have a lower
limit because of the gauge-link direction $v$. The $k^-$
components also have a lower limit: when the soft gluons are
factored, the gluons with $k^-/k^+$ smaller than $\tilde
v^-/\tilde v^+$ have been factored out of the jet.

\begin{figure}
\begin{center} \begin{picture}(330,70)(0,0) \SetOffset(20,10)
\SetWidth{0.7} \ArrowLine(10,10)(10,56)\ArrowLine(70,56)(70,10)
\Line(10,56)(32,56)\Line(10,54)(32,54)\Line(32,54)(32,56)
\Line(70,56)(48,56)\Line(70,54)(48,54)\Line(48,54)(48,56)
\Gluon(25,10)(25,56){-2}{6}\Gluon(18,10)(18,56){-2}{6}
\Gluon(55,10)(55,56){2}{6}\Gluon(63,10)(63,56){2}{6}
\GOval(40,10)(34,6)(90){0.8}\Text(40,10)[c]{$J$}\DashLine(40,70)(40,-10){5}
\SetWidth{1.8}\ArrowLine(-12,0)(6,10) \ArrowLine(74,10)(92,0)
\Text(-14,0)[r]{$P$}\Text(94,0)[l]{$P$} \Text(110,30)[c]{$=$}
\SetOffset(155,10)\SetWidth{0.7}
\ArrowLine(10,10)(10,56)\ArrowLine(70,56)(70,10)
\Line(10,56)(32,56)\Line(10,54)(32,54)\Line(32,54)(32,56)
\Line(70,56)(48,56)\Line(70,54)(48,54)\Line(48,54)(48,56)
\Gluon(25,10)(25,56){-2}{6}\Gluon(18,10)(18,56){-2}{6}
\Gluon(55,10)(55,56){2}{6}\Gluon(63,10)(63,56){2}{6}
\GOval(40,10)(34,6)(90){0.3}\Text(40,10)[c]{$J$}
\DashLine(40,70)(40,-10){5} \SetWidth{1.8}\ArrowLine(-12,0)(6,10)
\ArrowLine(74,10)(92,0) \SetWidth{0.7} \Line(100,-10)(126,65)
\SetOffset(250,-18) \COval(70,55)(11,7)(90){Black}{Red}
\SetColor{Red}
\Gluon(52,39)(65,50){1.5}{3}\Gluon(88,39)(75,50){1.5}{3}
\Gluon(52,71)(65,60){1.5}{3}\Gluon(88,71)(75,60){1.5}{3}
\SetColor{Black} \Line(40,55)(62,25)\Line(43,55)(65,25)
\Line(40,55)(62,85)\Line(43,55)(65,85)
\Line(100,55)(78,25)\Line(97,55)(75,25)
\Line(100,55)(78,85)\Line(97,55)(75,85)
\Text(70,55)[c]{$S$}\end{picture} \end{center} \caption{\it
Relation between soft-subtracted (left-hand side) and
un-subtracted parton distributions. The subtracted distribution
has a light-gray blob and the un-subtracted one has a dark blob.
The denominator on the right-hand side is a soft factor.}
\end{figure}
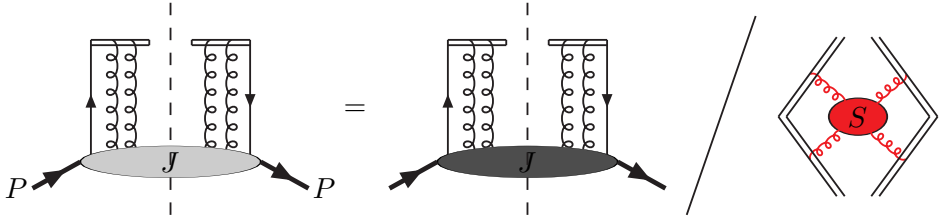

Therefore the jet factor is not the same as the parton
distribution ${\cal Q}(x, b, \mu^2, x\zeta)$ defined in Sec. II.
Rather, it is the same as the soft-subtracted parton distribution
$q(x,b, \mu^2, x\zeta, \rho)$. The relationship of the two is
shown in Fig. 9.

The renormalization group equations for ${\cal Q}$ and $\hat{\cal
Q}$ are known. After subtracting the soft factor, the equation has
to be modified by including the anomalous dimension for the soft
factor:
\begin{equation}
    \mu\frac{dq(x,b,\mu^2,x\zeta, \rho)}{d\mu} = \left(2\gamma_F -\gamma_S(\rho)\right)q(x,
    b,\mu^2,x\zeta, \rho)\ .
\end{equation}
A similar equation holds for the TMD fragmentation function. The
Collins and Soper equation for $q$ is the same as for ${\cal Q}$:
\begin{equation}
    \zeta\frac{\partial}{\partial \zeta}q(x,b,\mu^2,x\zeta, \rho)
      = \big(K(\mu, b)+G(\mu, x\zeta)\big)q(x,b,\mu^2,x\zeta, \rho) \
      .
\end{equation}
We will solve this equation to sum over large logarithms in Sec.
V.

\subsection{Subtraction Method}

For a given Feynman diagram, there are multiple leading regions.
To ensure the factorization works, one, in principle, has to
supply a subtraction method which allows separating contributions
from different leading regions. In particular, the subtraction
method must provide a systematic way of handling the overlapping
contribution of different leading regions.

The easiest way to develop a subtraction method in gauge theory is
to choose the axial gauge. For example, the factorization for
inclusive DIS in the axial gauge can be developed using the
Bethe-Salpeter formalism and has been used to calculate the
anomalous dimension of parton distributions at two-loop order
\cite{CurFurPet80}. For factorization involving collinear and soft
divergences, a subtraction method in the axial gauge has been
developed by Collins and Soper \cite{ColSop81}.

In covariant gauge, a systematic subtraction is complicated, and
has not yet been fully developed in the literature. At the
one-loop level, an example has been provided by Collins and
Hautmann \cite{ColHau00}. We have checked that the new subtraction
method corresponds to a particular choice of $\rho$ in this paper.
It would be interesting to pursue this subtraction to
higher-order. In particular, higher-order calculations help to
clarify the roles of different non-perturbative matrix elements
which have the same one-loop result.

The best approach to treating overlapping infrared divergences in
a multi-loop case might be the soft-collinear effective theory
mentioned in the introduction. Here we assume this can be done in
principle and leave a more careful discussion for future
publication.

\subsection{Factorization and $\rho$-Independence}

Collecting all factors in Fig. 8, one finally has the following
factorization formula:
\begin{eqnarray}
   F(x_B,z_h,b, Q^2) &=& \sum_{q=u,d,s,...} e_q^2
      q\left(x_B, z_hb, \mu^2, x_B\zeta, \rho\right)\hat q\left(z_h,b,\mu^2,
   \hat \zeta/z_h, \rho\right) \nonumber \\
   && \times S(b,\mu^2, \rho) H\left(Q^2,\mu^2,\rho \right)\ ,
\end{eqnarray}
advertised earlier.

It can be shown that the above expression is independent of
$\rho$. At one-loop order, this is easy to see,
\begin{equation}
     \rho \frac{\partial S(b,\rho)}{\partial \rho}
      = -\frac{\alpha_s C_F}{\pi} \ln\left(
        \frac{\mu^2b^2}{4} e^{2\gamma_E}\right) S(b, \rho) \ ,
\end{equation}
On the other hand,
\begin{equation}
   \rho \frac{\partial S(b,\rho)q(x,b,\rho)}{\partial \rho}
      = -\frac{\alpha_s C_F}{2\pi} \ln\left(
        \frac{\rho Q^2b^2}{4} e^{2\gamma_E-1}\right)S(b, \rho) q(x, b,
        \rho)\ .
\end{equation}
If the structure function is independent of $\rho$, the above
requires $H(\rho)$ to evolve in $\rho$,
\begin{equation}
   \rho \frac{\partial H(\rho)}{\partial \rho}
      = \frac{\alpha_s C_F}{\pi} \left[\ln\left(
        \frac{\rho Q^2}{\mu^2}\right)-1\right] H(\rho) \ .
\end{equation}
It is easy to check that our one-loop hard part satisfies the
above equation.

At higher orders, the $\rho$ independence is guaranteed because
one can view the factorization formula as a definition for the
hard part.

\section{Summing over large Logarithms}

From the factorization formula and the evolution equations, we can
get an expression for the structure function in which the large
logarithms involving momentum $Q$ are summed over.

First of all, there are large logarithms in $K+G$ (which is
independent of the renormalization scale). To sum it, we solve the
renormalization group equation to get
\begin{equation}
     K(b,\mu) + G(x\zeta, \mu)
      = K(b,\mu_L) + G(x\zeta, \mu_H) - \int^{\mu_H}_{\mu_L}
      \frac{d\tilde \mu}{\tilde \mu} \gamma_K(\alpha(\tilde\mu)) \
      .
\end{equation}
To isolate the large logarithms, one has to choose $\mu_L$ to be
on the order of $\Lambda_{\rm QCD}$ and $\mu_H$ to be on the order
of $\zeta$. Therefore, we let
\begin{equation}
    \mu_L= C_1 M_N; ~~~\mu_H = C_2x\zeta =C_2 Q\sqrt{\rho}\ ,
\end{equation}
where $M_N$ is the mass of the nucleon.

Substituting the above into the Collins-Soper equation for $q(x,
b, \mu^2, x\zeta, \rho)$, the large logarithms in $\zeta$ can be
factorized,
\begin{eqnarray}
     q(x,b,\mu,x\zeta,\rho) &=& \exp\left\{-\int^{C_2x\zeta}_{\mu_L} \frac{d\mu}{\mu}
         \left[\ln\left(\frac{C_2x\zeta}{\mu}\right)\gamma_K(\alpha(\mu))
          - K(b,\mu_L) - G(\mu/C_2, \mu)\right]\right\}
     \nonumber \\
    && \times q(x,
          b,\mu,x\zeta_0=\mu_L/C_2, \rho) \ ,
\end{eqnarray}
where the exponential factor contains the entire dependence on
$\zeta$, in particular, the large Sudakov double logarithms.
However, the above expression contains much more than just the
leading double logarithms; it contains all the sub-leading logs as
well.

Similarly, one can find the solution for the fragmentation
function,
\begin{eqnarray}
     \hat q(z,b,\mu,\hat\zeta/z,\rho) &=& \exp\left\{-\int^{C_2\hat\zeta/z}_{\mu_L} \frac{d\mu}{\mu}
         \left[\ln\left(\frac{C_2\hat \zeta}{z\mu}\right)\gamma_K(\alpha(\mu))
          - K(b,\mu_L) - G(\mu/C_2, \mu)\right]\right\}
     \nonumber \\
    && \times \hat q(z,
          b,\mu,\hat \zeta_0/z=\mu_L/C_2, \rho) \ .
\end{eqnarray}
If we choose a frame in which $x_B\zeta = \hat \zeta/z_h$, then
the exponential factors in $q$ and $\hat q$ become the same, and
moreover $\zeta^2x_B^2  = \hat \zeta^2/z_h = Q^2\rho $.

Let us study the renormalization group equation for the hard part.
The physical cross section is, of course, independent of the
renormalization scale $\mu$. Since we know the renormalization
group equation for $q$, $\hat q$ and the soft factor, we can
easily derive the renormalization group equation for the hard
part,
\begin{equation}
  \mu\frac{dH(Q^2/\mu^2,\rho)}{d\mu} =
  -\left(4\gamma_F -\gamma_S(\rho)\right)H(Q^2/\mu^2,\rho) \ .
\end{equation}
The solution is
\begin{equation}
     H(Q^2/\mu^2,\rho) = \exp\left\{ -\int^\mu_{\mu'} \frac{d\mu}{\mu}
     [4\gamma_F - \gamma_S(\rho)]\right\} H(Q^2/\mu^{'2},\rho) \ .
\end{equation}
To factor out the large renormalization logarithms, one can choose
$\mu$ to be at low scale such as $\mu_L$, and $\mu'$ at high scale
such as $\mu_H$. Therefore, we write,
\begin{equation}
       H(Q^2/\mu_L^2,\rho) = \exp\left\{ -\int^{\mu_L}_{C_2x\zeta} \frac{d\mu}{\mu}
     [4\gamma_F - \gamma_S]\right\} H(Q^2/C_2x\zeta, \rho) \ ,
\end{equation}
where $H(Q^2/C_2x\zeta, \rho)$ contains no large logarithms.

Collecting the above results, one has both the renormalization and
soft-collinear logarithms summed in the following expression:
\begin{eqnarray}
   F(x_B,z_h,b, Q^2) &=&
      q\left(x_B, z_hb,\mu_L^2, \mu_L/C_2, \rho\right)\hat q\left(z_h,b,\mu_L^2,
   \mu_L/C_2, \rho\right) S(b,\mu_L^2, \rho) H\left(1/C_2^2\rho,\rho\right) \nonumber  \\
&& \times \exp\left\{-2\int^{C_2Q\sqrt{\rho}}_{\mu_L}
\frac{d\mu}{\mu}
         \left[\ln\left(\frac{C_2Q\sqrt{\rho}}{\mu}\right)\gamma_K(\alpha(\mu))
          - K(b,\mu_L) \right.\right. \nonumber \\ &&
         ~~~~~~~~ \left.\left. - G(\mu/C_2, \mu) - 2 \gamma_F+\frac{1}{2}\gamma_S(\rho)\right]\right\} \
         ,
\end{eqnarray}
where all large logarithms have been factorized in the exponential
factor. For the physics discussed in this paper, we do not want
significant large logarithms, because otherwise the transverse
momentum of the hadron yield is generated mostly by soft-gluon
radiations. To avoid them, $Q^2$ can only be moderately large
compared to $P_{h\perp}$. On the other hand, in this kinematic
regime, the contributions from power-suppressed terms might not be
entirely negligible.

Finally, the choice of $\rho$. According to its definition, we
must have $\rho\gg 1$ although the physics is independent of
$\rho$.  However, if $\rho$ is too large, one has large logarithms
in the hard part and the convergence of the perturbation series
might be spoiled. Therefore in practice one might choose a $\rho$,
for example, somewhere in between 3 and 10.

\section{Conclusion}

In this paper, we showed that a factorization theorem exists for
semi-inclusive deep-inelastic scattering with detected hadron
momentum $P_{\perp h}\ll Q$. $P_{h\perp}$ can either be soft,
i.e., on the order of $\Lambda_{\rm QCD}$, or in the perturbative
domain $\gg \Lambda_{\rm QCD}$. We have mainly focused on the
former case although the result is valid also for the latter. For
$P_{\perp h}\gg \Lambda_{\rm QCD}$, the theorem can be simplified
by an additional factorization of the TMD parton distributions and
fragmentation functions \cite{ColSop81}.

We established the theorem by first considering the example at the
one-loop level. In this case, the calculation of the parton
distribution, fragmentation function, soft factor, and the SIDIS
cross section was straightforward. The example demonstrated that
the factorization indeed works.

At higher-order in perturbation theory, one can use the formalism
developed by Collins, Sterman, and Soper and others. Starting from
the most general reduced diagrams, we showed the factorization of
collinear gluons from the hard part and the soft gluons from the
collinear part, matching the jet factors with the distribution and
fragmentation functions. The factorization scale $\mu$ and the
$\rho$ independence of the physical cross section allows one to
sum over large logarithms involving scales $Q$ and $P_{h\perp}$.

The present result can be easily extended to the situations where
the target is polarized or the polarization of the final state
hadron is measured. It can also be extended to the case where the
transverse momentum of the hadrons are integrated with a weighting
factor. Finally, all results here can be obtained also in the
framework of the soft-collinear effective theory.

We would like to thank J. Collins for useful discussions related
to the subject of this paper. X. J. and F. Y. were supported by
the U. S. Department of Energy via grant DE-FG02-93ER-40762.
J.P.M. was supported by National Natural Science Foundation of
P.R. China (NSFC). X. J. is also supported by an Overseas
Outstanding Young Chinese Scientist grant from NSFC.

\end{document}